\documentclass[%
reprint,
superscriptaddress,
amsmath,amssymb,
aps,
prb,
]{revtex4-2}

\usepackage{booktabs}
\usepackage{multirow}
\usepackage{color}
\usepackage{graphicx}
\usepackage{dcolumn}
\usepackage{bm}
\usepackage{hyperref}
\usepackage[mathlines]{lineno}
\usepackage{siunitx}
\usepackage{amsmath}

\begin{document}
	
	\title{Designing Guidance for Multiple Valley-based Topological States Driven by Magnetic Substrates: Potential Applications at High Temperatures}

\author{Xiyu Hong}
\affiliation{%
	State Key Laboratory of Low Dimensional Quantum Physics, Department of Physics, Tsinghua University, Beijing 100084, China
}%

\author{Zhe Li}
\email{lizhe21@iphy.ac.cn}
\affiliation{Beijing National Research Center for Condensed Matter Physics, and Institute of Physics, Chinese Academy of Sciences, Beijing 100190, China}

	\date{\today}
	
	\begin{abstract}
Valley-based topological phases offer a wealth of exotic quantum states with tunable functionalities, driven by the valley degree of freedom. In this work, by constructing heterostructures of germanene (silicene, stanene) on various magnetic substrates, we address key tuning factors such as the spin-orbit coupling (SOC) strength of the substrate, magnetic orientations, and stacking orders, all of which govern multiple valley-based topological features. We present a comprehensive guiding principle for the efficient manipulation of these features, achieved simply by designing and modulating the magnetic properties of the underlying substrates. Specifically, increasing the SOC strength of the magnetic substrate facilitates a range of topological phase transitions characterized by different Chern numbers, with many systems exhibiting a transition from quantum valley Hall to quantum anomalous Hall (QAH) states. Additionally, rotating the in-plane magnetic orientation of the substrate enables tunability of the Chern number and chirality, within a moderate range of SOC strength. Furthermore, the antiferromagnetic coupling of the magnetic substrate can induce valley-based QAH states with substantial valley gaps, leveraging its high Curie temperature ($T_{\rm C}$) to enable the realization of multiple tunable magnetic topologies at elevated temperatures. Our findings provide a straightforward strategy for the design and manipulation of spintronic and valleytronic devices that can potentially operate under high-temperature conditions.
	\end{abstract}
	
	\maketitle

\section{Introduction}

The quantum anomalous Hall effect (QAHE), characterized by dissipationless edge states and quantized Hall conductance in the absence of an external magnetic field, stands as a cornerstone in modern condensed matter physics\cite{haldane1988model,yu2010quantized,hasan2010colloquium,qi2011topological,weng2015quantum,chang2023colloquium,wang2023intrinsic}. Since its first experimental realization in magnetically doped topological insulators (TIs), such as Cr-doped thin films \cite{chang2013experimental,mogi2015magnetic,ou2018enhancing}, QAHE has provided unprecedented opportunities for exploring dissipationless electronic currents, Majorana fermions, and topological magnetoelectric effects\cite{hasan2010colloquium,qi2011topological}. Majorly, there exist several distinct designing strategies supporting the realization of QAHE, involving the aforementioned magnetically doped TIs \cite{chang2013experimental,mogi2015magnetic,ou2018enhancing}, magnetic-insulator/topological-insulator proximity systems \cite{otrokov2017highly,grutter2021magnetic,tss-li2024realization,li2024universally,cgt-mogi2019large,cgt-yao2019record,cgt-alegria2014large,EuS-wei2013exchange,MnTe-he2018exchange,ZnCrTe-watanabe2019quantum,xue2024tunable}, and intrinsic magnetic topological insulators like MnBi$_2$Te$_4$-family materials \cite{li2019intrinsic,otrokov2019unique,zhang2019topological,gong2019experimental,deng2020quantum,liu2020robust,ge2020high,bai2024quantized,li2020tunable,kobialka2022dynamical,wang2023magnetic,guo2023novel,niu2019quantum,zhang2020large,eremeev2022magnetic,tang2023intrinsic,yang2024exploration,wu2023high}. Within these building strategies, the inner magnetisms include $p$-$d$ hybridization that exchange mass terms induce in the low-energy Dirac-like Hamiltonian (predicted and observed in Cr-doped TIs, MnBi$_2$Te$_4$, etc.), and $d$-$d$ correlation mechanism that predicted in layered magnets like Pd(Pt)Br(I)$_3$, MnBr$_3$, LiFeSe, Pd(Ni)Sb(As, Bi)O$_3$, V$_2$$MX_4$ \cite{you2019two,li2023tunable,li2020high,li2022chern,wu2023robust,jiang2024monolayer}, and so on.

Above all, importing valley degree of freedom enriches the tunability of quantum anomalous Hall conductance and application prospect in future manipulation of multiple valley-based topology. Recently, comprehending and designing multifunctional valley-based topological states also attracts extensive attention \cite{pan2014valley,pan2015valley,zhang2015robust,zhang2018strong,zou2020intrinsic,vila2021valley,zhou2017valley,li2024multimechanism,xue2024valley}. Specially, profited by its prominent characters involving two-dimensional (2D) configuration, suspended $p_z$ orbitals and the moderate strength of spin-orbit coupling (SOC), germanene, silicene, and stanene play a vital role in fabricating valleytronic devices and further realization and investigation of novel valley-based topological properties \cite{li2024multimechanism,xue2024valley}. Remarkably, the magnetic substrate furnishes an exotic modulable device towards abundant topological phases containing not only normal QAHE, but also valley-polarized QAHE (vpQAHE) \cite{li2024multimechanism,xue2024valley}, multimechanism QAHE \cite{li2024multimechanism}, quantum valley Hall effect (QVHE) \cite{xue2024valley}, quantum spin Hall effect (QSHE) \cite{xue2024valley} without time reversal symmetry (TRS) via selecting typical 2D magnets involving Cr$_2$Ge$_2$Te$_6$ (CGT) and $X$Bi$_2$Te$_4$ ($X$ = Mn, Ni, V), most of which fit well with germanene-family films on the in-plane lattice constants \cite{li2024multimechanism,xue2024valley}. Concretely, the magnetic substrate offers the Zeeman splitting term from the magnetism, and the external Rashba SOC term originating from the breaking of inversion symmetry and the SOC strength of the substrate \cite{pan2014valley,pan2015valley}. Consequently, the immediate design of valley-based topologies is to manipulate the magnetic configurations and orientations, and to aptly vary the SOC strength by choosing different elements as components in the substrate.

The discovery and successful preparation of layered ferromagnetic (FM) insulators that maintain their magnetism down to monolayer trigger enormous investigations into their outstanding potential for multifunctional substrates. For instance, CrI$_3$ and CGT yield both high Curie temperature and strong magnetic crystalline anisotropy energies (MAEs) with out-of-plane magnetism \cite{huang2017layer,gong2017discovery,sivadas2018stacking}; MnBi$_2$Te$_4$-family materials possess both the monolayer-based FM state and topology which manifest them as intrinsic magnetic Chern insulators\cite{li2019intrinsic,zhang2019topological,otrokov2019unique,li2020tunable}. These magnetic building blocks supply plentiful modulating degrees. Further intercalation of $X$Te-layer ($X$ = Mn, Ni, V) into MnBi$_2$Te$_4$-family materials \cite{zhang2020large,eremeev2022magnetic,tang2023intrinsic,yang2024exploration,wu2023high} even adds more degrees of freedom into valley-based topological manipulations with bi-magnetic-layer-induced critical temperature promotions. By dint of the aforementioned 2D magnets with germanene (silicene, stanene), especially $XY$Bi$_2$Te$_5$ ($X$, $Y$ = Mn, Ni, V), which include higher executable tunability, it’s facilitated to establish a universal design guidance for valley-based topological phase diagrams simply by modifying the magnet underneath, a guidance which is currently still lacking.

In this work, via establishing germanene (silicene, stanene)/$XY$Bi$_2$Te$_5$ and comparing it with germanene/$X$Bi$_2$Te$_4$ ($X$ = Mn, Ni), germanene/NiAs-Mn$_3$Bi$_2$Te$_6$ and germanene/CGT, involving manipulating parameters such as the magnetic substrate-SOC strengths, the magnetic moment orientations, stacking-orders, and inter-van-der-Waals (vdW) spacings, we provide a universal guidance towards how to design and manipulate multiple valley-based topological characters. Our findings show that by increasing the substrate-SOC strength, the germanene grown above experiences multiple phase transition points characterized by various Chern numbers in between. In many cases, a QVHE-QAHE phase transition appears. Rotating in-plane magnetic orientations successfully drives both the chirality and the value of Chern number manipulations within a moderate range of substrate-SOC strengths. Stacking-order shifts serve as an experimentally executable way to tune the value and chirality of Chern numbers among Ni-neared building-blocks. Moreover, the antiferromagnetic (AFM) coupling nature of magnetic substrates can also serve to induce the valley-based QAHE in germanene while exploiting its high critical temperature ($T_{\rm C}$) character with sufficient valley gaps (for example, the vpQAHE state in germanene/NiAs-NiMnBi$_2$Te$_5$ with $T_{\rm C}$ as 152K and $K$ valley gap as 10.9meV), facilitating the prediction and investigation of high-temperature, valley-based multiple and modulable magnetic topology.

 \label{i:introduction}

\section{Methods}

We employ first-principles computational approaches using the Vienna Ab Initio Simulation Package (VASP)\cite{kresse1996efficient,perdew1996generalized}, in combination with tight-binding Hamiltonian and Green's function methods. Structural relaxations, static self-consistent computations, and band structure calculations were conducted on an $11 \times 11 \times 1$ Monkhorst-Pack $k$-point grid\cite{monkhorst1976special}. During relaxation, atomic geometries were optimized until Hellmann-Feynman forces on all atoms fell below 0.0031 eV/Å, with an energy convergence criterion of $1.0 \times 10^{-7}$ eV.

For correlated $3d$ orbitals in magnetic elements, we implemented the PBE+$U$ \cite{perdew1996generalized} functional with $U$ values set to 3 eV for V 3$d$, 4 eV for Mn 3$d$, and 4 eV for Ni 3$d$ orbitals \cite{li2019intrinsic,li2020tunable,tang2023intrinsic,li2024multimechanism,li2024universally}. Structural relaxation steps employed collinear magnetic moments without spin-orbit coupling (SOC), while noncollinear magnetic moments with SOC were included for self-consistent and band structure calculations to fully capture topological effects. vdW corrections via the DFT-D3 method\cite{grimme2010consistent} were utilized to account for interlayer coupling, with a 20 Å vacuum layer added to prevent interactions between periodic images\cite{tkatchenko2009accurate}. VASPKIT is used for post-processing of band structures \cite{wang2021vaspkit}.

To evaluate interlayer exchange coupling in $XY$Bi$_2$Te$_5$ substrates, the energy difference between AFM and FM configurations was calculated by employing noncollinear magnetic moments with SOC. MAE was determined by computing the energy difference between in-plane ($y$-axis) and out-of-plane ($z$-axis) magnetization, considering isotropic in-plane spin directions, also under noncollinear magnetic moment configurations with SOC. Stacking-order shifts in the heterostructures were systematically analyzed using $6 \times 6$ in-plane meshes.

For topological properties such as edge states, Berry curvature distributions, and Chern numbers, tight-binding Hamiltonians were constructed using maximally localized Wannier functions via the Wannier90 package\cite{marzari1997maximally,souza2001maximally,mostofi2014updated}. Green's function techniques from WannierTools\cite{wu2018wanniertools} were employed to extract multiple topological features based on the obtained tight-binding Hamiltonians. Additionally, to precisely extract the local gap values around each valley, we utilized plane-band computations by casting $401 \times 401$ meshes on 2.56\% of the area of the first 2D Brillouin zone centered at the $K'$ or $K$ point, based on the obtained tight-binding Hamiltonians.

Furthermore, Monte Carlo (MC) simulations were employed with the Heisenberg spin model to obtain the $T_{\rm C}$ of long-range magnetism, executed by the package \textit{mcsolver} \cite{liu2019magnetic}. Specifically, in the MC simulations, $16 \times 16 \times 1$ superlattices with more than $3 \times 10^5$ and $6 \times 10^5$ steps of the Metropolis algorithm were adopted to reach thermal equilibrium if the $T_{\rm C}$ value is below or above 100K, respectively. The 1K (2K) step size was chosen to precisely acquire the value of $T_{\rm C}$ when the $T_{\rm C}$ value is below (above) 300K.
\begin{center}
	\begin{figure*}
		\centering
		\includegraphics[width=1\linewidth]{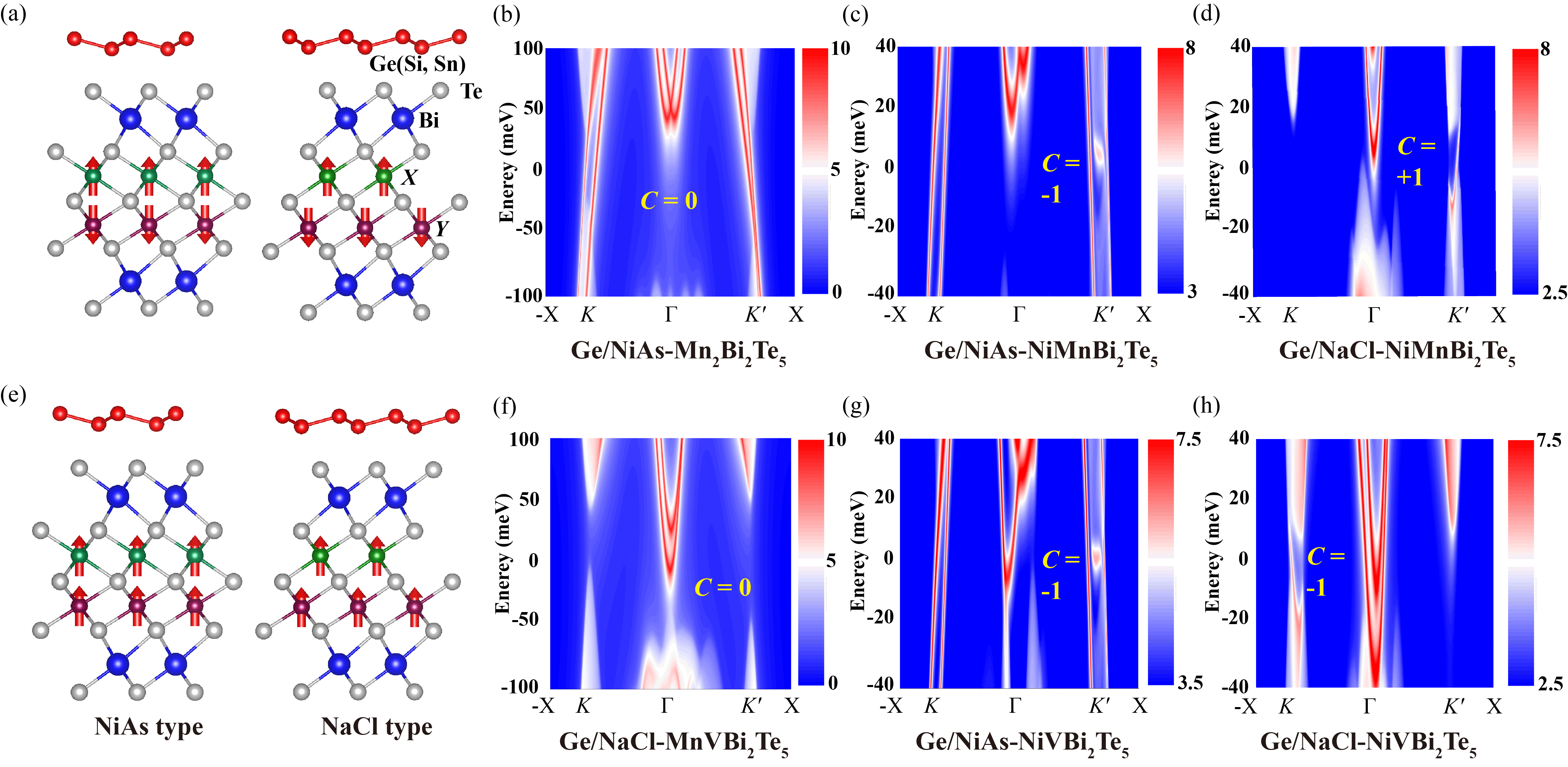}
		\caption{Side-view structures and local-density-of-state (LDOS) patterns of Ge/$XY$Bi$_2$Te$_5$ systems.
		(a) and (e) depict the two typical structures (i.e. NiAs type and NaCl type) of Ge/$XY$Bi$_2$Te$_5$ in which 
		the bottom $XY$Bi$_2$Te$_5$ layers contain AFM and FM interatomic-layer couplings respectively. Red, purple, green, blue and light-gray balls correspond to Ge, $Y$, $X$, Bi, and Te atoms respectively, while the red vectors represent the directions of the magnetic moments. (b)-(d) provide the LDOS patterns of germanene grown on AFM substrate cases: NiAs-Mn$_2$Bi$_2$Te$_5$, NiAs-NiMnBi$_2$Te$_5$ and NaCl-NiMnBi$_2$Te$_5$. Similarly, (f)-(h) reveal the LDOS patterns under FM substrate cases: NaCl-MnVBi$_2$Te$_5$, NiAs-NiVBi$_2$Te$_5$, and NaCl-NiVBi$_2$Te$_5$. Among all the LDOS patterns, the value of LDOS rises from blue to white and then to red colors.}
		\label{fig:figure1}
	\end{figure*}
\end{center}

	Finally, in order to confirm the structural stability, we adopted density functional perturbation theory (DFPT) to calculate phonon spectra without implementing supercell lattice (i. e. setting the $1\times1\times1$ supercell) \cite{togo2015first}. These calculations were performed under collinear spin configurations without SOC to evaluate dynamical stability. The package of PHONOPY \cite{togo2023first} is employed as the post-processing tool to obtain the final phonon spectrums.

	 \label{ii:method}
	
\section{Basic Structures and Band Topology}

Xene refers to a class of buckled two-dimensional (2D) materials with a honeycomb lattice structure \cite{molle2017buckled}, including carbon-family materials like germanene, silicene, and stanene. Unlike graphene, where all carbon atoms reside within a single plane, Xene exhibits a buckled geometry due to its $sp^3$-like hybridization, resulting in a bilayer (111) configuration where alternating atoms are displaced vertically \cite{vogt2012silicene,pan2014valley,xu2013large,xu2015large,zhu2015epitaxial}. This unique structure enhances inversion symmetry breaking and facilitates hybridization with substrate materials. Moreover, the relatively wide range of lattice constants in Xene allows strain engineering and efficient interfacing with a variety of magnetic substrates, making it an ideal platform for designing heterostructures with tunable topological properties \cite{davila2014germanene,fleurence2012experimental}.

In this study, considering the plentful tunability, $XY$Bi$_2$Te$_5$ ($X$ = Mn, Ni, $Y$ = Mn, Ni, V) family materials are majorly chosen as the substrate for exploring the interplay between magnetic and topological properties. $XY$Bi$_2$Te$_5$ belongs to a family of van der Waals (vdW) layered materials, derived from Mn$_2$Bi$_2$Te$_5$ by substituting Mn atoms with other magnetic elements \cite{tang2023intrinsic}. The lattice structure consists of a nine-atomic-layer stacking sequence: Te–Bi–Te–$X$–Te–$Y$–Te–Bi–Te, where $X$ and $Y$ represent magnetic atoms. This structure introduces additional control over interlayer coupling, magnetic anisotropy, and SOC strength, offering significant flexibility for tuning topological phases compared to simpler systems like MnBi$_2$Te$_4$ \cite{tang2023intrinsic, li2019intrinsic,zhang2019topological,otrokov2019unique,gong2019experimental,deng2020quantum,liu2019magnetic,yu2010quantized,bai2024quantized,li2020tunable}.

The magnetic and topological properties of $XY$Bi$_2$Te$_5$ are closely related to its layered structure and stacking configuration. Previous studies have shown that $XY$Bi$_2$Te$_5$ can adopt two distinct atomic stacking configurations: NiAs type ("$ABAC$" stacking) or NaCl type ("$ABC$" stacking), depending on the combination of $X$ and $Y$ elements \cite{eremeev2022magnetic,tang2023intrinsic,yang2024exploration}. Moreover, comparing to traditional NaCl type structures, NiAs type stacking enhances interatomic interactions, promoting strong magnetic exchange coupling, dedicates more novel types of magnetic topological characters. Performing as the basic structures of the beneath substrates and devoting the magnetism, these stacking configurations are critical for modulating the Berry curvature distribution and valley-specific topological properties of the whole building blocks. \cite{li2024multimechanism,li2023stacking,zhao2023stacking}.

Phonon spectrum calculations confirm the dynamical stability of these heterostructures, with the results depicted in Figs. S1-S4 of the Supplementary Materials \cite{supplementary}. Obviously no imaginary frequency exists in all the heterostructures mentioned in this work [Fig. S1] expect the case germanene (Ge)/NaCl-Ni$_2$Bi$_2$Te$_5$ that the whole Brillouin-zone (BZ) distributed imaginary frequency emerges, indicating the structrual instability. Hence, we ignore it in the following investigations and discussions. Furthermore, considering the uncertainty of in-plane lattice constants that may be brought by experimental fabrications, we also make structural stability trials of Xenes and magnetic substrates under biaxial strains, with the outcomes summarized in Figs. S2-S4. Within $\pm 5\%$ of the biaxial strains, most building ingredients maintains the stability, except for NaCl-Mn$_2$Bi$_2$Te$_5$ (above +2.5\%), NaCl-MnNiBi$_2$Te$_5$ (above +1.0\%) and NaCl-NiVBi$_2$Te$_5$ (above +3.0\%) that generates damages via implementing certain lattice stretches.

By leveraging the tunable structural properties of $XY$Bi$_2$Te$_5$ substrates and their compatibility with the buckled structure of germanene, this study systematically explores the topological properties of Xene/$XY$Bi$_2$Te$_5$ heterostructures, comparing with Ge/$X$Bi$_2$Te$_4$, Ge/NiAs-Mn$_3$Bi$_2$Te$_6$ and Ge/CGT to extract the universally applicable guidance. It may be a little bit far from freely and experimentally producting various candidates of $XY$Bi$_2$Te$_5$ (up to now Mn$_2$Bi$_2$Te$_5$ and Mn$_2$Sb$_2$Te$_5$ has been fabricated \cite{cao2021growth,saxena2023growth}), but the highly tunability of $XY$Bi$_2$Te$_5$-family magnetic substrates offer a focal reference for selections of other analogous magnetic substrates that sharing the similar properties. In the following discussions, by default we use the notation "Xene/$XY$Bi$_2$Te$_5$" in which "$X$" and "$Y$" stand for the nearest and second nearest magnetic atomic-layer to Xene-layer respectively.

\begin{center}
	\begin{figure*}
		\centering
		\includegraphics[width=1\linewidth]{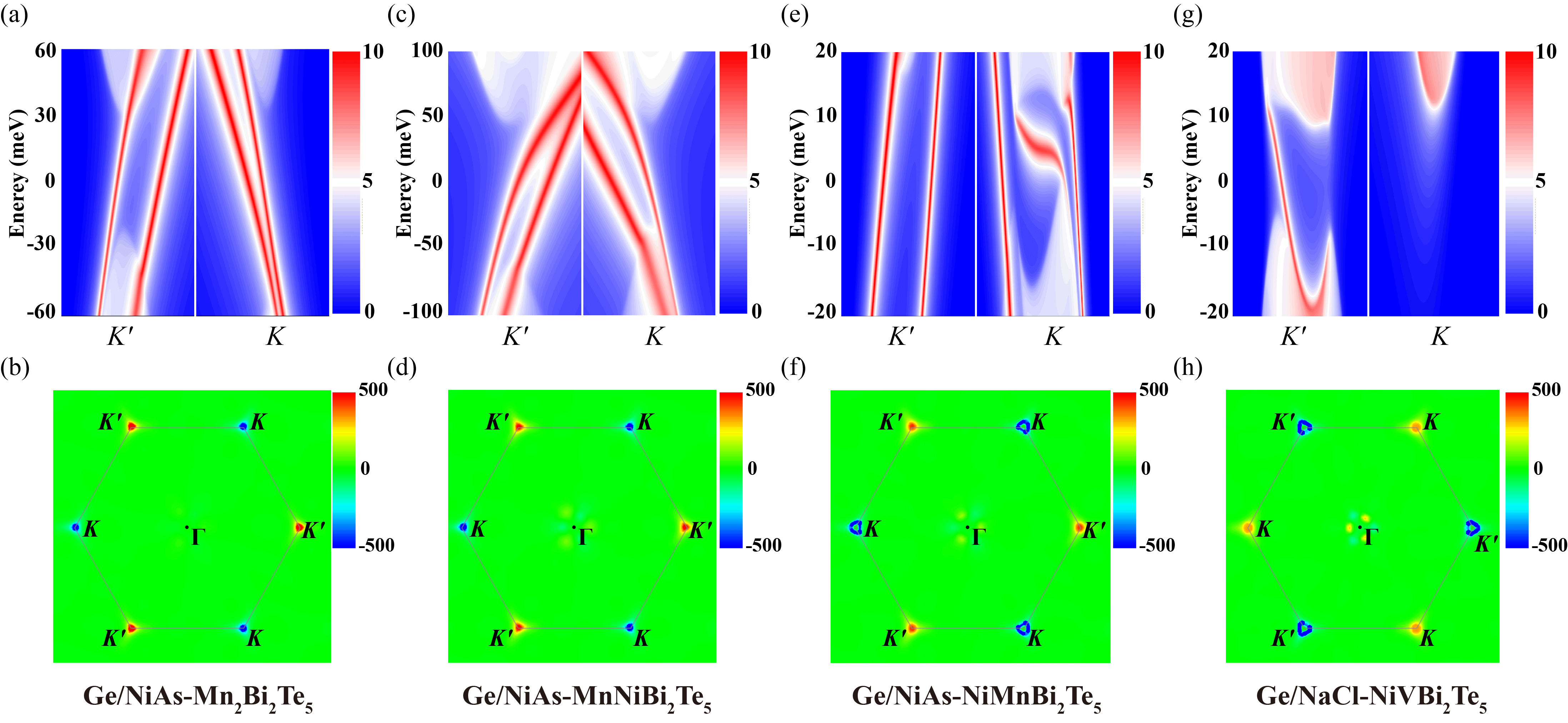}
		\caption{Valley-adjacent zoom-in LDOS patterns and the Berry curvature distributions of the four representative Ge/$XY$Bi$_2$Te$_5$ systems. (a) Zoom-in LDOS patterns of Ge/NiAs-Mn$_2$Bi$_2$Te$_5$ near the $K’$ and $K$ valley gaps, separated with a white vertical cut-off line. (b) Berry-curvature-distributed mapping of Ge/NiAs-Mn$_2$Bi$_2$Te$_5$ around a little bit more than one BZ, within which the blue and red colors are related to the negative and positive chirality of Berry curvature respectively, while the most-distributed green zone verifies the zero value of Berry curvature. (c), (e) and (g) share the similar valley-adjacent zoom-in LDOS patterns, but under the magnetic substrate as NiAs-MnNiBi$_2$Te$_5$, NiAs-NiMnBi$_2$Te$_5$ and NaCl-NiVBi$_2$Te$_5$ accordingly. Analogously, (d), (f) and (h) draw the Berry curvature distributions of Ge/NiAs-MnNiBi$_2$Te$_5$, Ge/NiAs-NiMnBi$_2$Te$_5$ and Ge/NaCl-NiVBi$_2$Te$_5$ respectively.}
		\label{fig:figure2}
	\end{figure*}
\end{center}

Figure~\ref{fig:figure1} illustrates the side view structures and local density of states (LDOS) patterns for six representative Ge/$XY$Bi$_2$Te$_5$ systems: involving AFM-ground-state Ge/NiAs-Mn$_2$Bi$_2$Te$_5$, Ge/NiAs-NiMnBi$_2$Te$_5$, Ge/NaCl-NiMnBi$_2$Te$_5$, and FM-ground-state Ge/NaCl-MnVBi$_2$Te$_5$, Ge/NiAs-NiVBi$_2$Te$_5$, Ge/NaCl-NiVBi$_2$Te$_5$. In Figs. \ref{fig:figure1} (a) and (e), the magnetic coupling (AFM or FM) within the substrate is shown with red arrows representing the magnetic moment directions. The LDOS patterns in Figure~\ref{fig:figure1}(b)-(d) and (f)-(h) reveal significant variations in the electronic density distributions near the $\Gamma$ and $K(K')$ points under different substrate configurations and magnetic couplings. Notably, the interplay between substrate symmetry, magnetic coupling and the mass terms induced from the nearest and the second nearest magnetic atomic-layer critically impacts the valley-specific topological features.

Similar to those of Ge/$X$Bi$_2$Te$_4$ \cite{li2024multimechanism,xue2024valley}, certain biaxial strains are also needed to be implemented on Ge/$XY$Bi$_2$Te$_5$ for the sake of the band alignment between the $\Gamma$ gap and the $K(K')$ valley gaps. Table \ref{tab1:biaxial} lists the concrete biaxial strains that forming the band-gap alignment in Ge/$XY$Bi$_2$Te$_5$ accompanied with Ge/NiAs-Mn$_3$Bi$_2$Te$_6$. Besides, the real in-plane lattice constants of these systems are enumerated in Table S1.

\begin{table}
	\caption{\label{tab1:biaxial} Biaxial strains that required to be implemented on Ge/$XY$Bi$_2$Te$_5$ and Ge/NiAs-Mn$_3$Bi$_2$Te$_6$ to reach the band gap alignment. Positive and negative ratios are corresponding to the stretch and compress strains that extracted by comparing to the free relaxed in-plane lattice constants.}
	\begin{ruledtabular}
		\begin{tabular}{cc}
			Building blocks & Ratio of biaxial strains        \\  
			\colrule
			Ge/NiAs-Mn$_2$Bi$_2$Te$_5$ &  0.0\% \\
			Ge/NaCl-Mn$_2$Bi$_2$Te$_5$ & +1.0\%  \\
			Ge/NiAs-MnNiBi$_2$Te$_5$ &  +2.0\% \\
			Ge/NaCl-MnNiBi$_2$Te$_5$ &  +1.5\%  \\
			Ge/NiAs-NiMnBi$_2$Te$_5$ &  +1.5\% \\
			Ge/NaCl-NiMnBi$_2$Te$_5$ &  +1.0\%  \\
			Ge/NiAs-MnVBi$_2$Te$_5$ &  0.0\% \\
			Ge/NaCl-MnVBi$_2$Te$_5$ &  0.0\%  \\
			Ge/NiAs-NiVBi$_2$Te$_5$ &  +2.0\% \\
			Ge/NaCl-NiVBi$_2$Te$_5$ &  +1.5\%  \\
			\colrule
		    Ge/NiAs-Mn$_3$Bi$_2$Te$_6$ &  +1.0\%  \\			
		\end{tabular}
	\end{ruledtabular}
\end{table}

In order to further dig out the valley-based topologies in the above systems, in Figure~\ref{fig:figure2} we provides a more detailed analysis of valley-resolved LDOS patterns and Berry curvature distributions for the four selected heterostructures: Ge/NiAs-Mn$_2$Bi$_2$Te$_5$, Ge/NiAs-MnNiBi$_2$Te$_5$, Ge/NiAs-NiMnBi$_2$Te$_5$ and Ge/NaCl-NiVBi$_2$Te$_5$. In the case of Ge/NiAs-Mn$_2$Bi$_2$Te$_5$, the inversion symmetry breaks at the interface, generating Berry curvature at both the $K$ and $K'$ valleys. However, the contributions from these valleys cancel each other, resulting in a trivial Chern number ($C = 0$) with QVHE. Analogously, Ge/NiAs-MnNiBi$_2$Te$_5$ also manifests it as QVHE state. However, the nearest Mn atomic-layer in the substrate replaced by Ni, i. e. taking Ge/NiAs-NiMnBi$_2$Te$_5$ as example, the balance between the $K$ and $K'$ valleys is much more disrupted mainly due to the larger mass terms induced from Ni compared to that of Mn \cite{xu2022controllable}. This leads to a vpQAHE state with asymmetric Chern numbers at the $K$ and $K'$ valleys and a total Chern number of $C = -1$. Ge/NaCl-NiVBi$_2$Te$_5$ shares the similar behavior (vpQAHE) with Ge/NiAs-NiMnBi$_2$Te$_5$, but with chirality-inversion of valley Chern number, stemming from the inversion of germanene-based atomic stacking-order compared to that of NiAs type cases, leading to the valley exchanges between $K$ and $K'$. All the systems above are handled with magnetic ground states.

Preiminarily, we can make a rough conclusion that only the nearest magnetic atomic-layer determines the final valley-based topology in Xene itself. Intuitively, it's easy to be comprehended that the mass terms induced by the magnetic superexchange mechanisms decay exponentially along with the increasing of the exchange distances. Nevertheless, for the case of FM-state Ge/NiAs-MnNiBi$_2$Te$_5$ QAHE state is successfully imported in germanene, distinct from AFM-ground-state case with QVHE state [Fig. S11(c)]. The contributions of the second nearest Ni layer in this substrate suffers little decadence benefitting from its NiAs type structure. For other systems by reversing the second atomic-layer magnetism, comparing Fig. S11 with Fig. \ref{fig:figure2}, no topological phase transition appears. Therefore, in most conditions the nearest-magnetic atomic-layer determination rule takes effect.

The above findings inspire us to exploit magnetic substrates (not limited to $XY$Bi$_2$Te$_5$-family materials) with AFM coupling nature to import valley-based QAHE states into Xene, only if the nearest magnetic atomic-layer retains FM state. This character facilitates us to select magnetic substrates with novel features among wider range of materials.

Overall, the band structure analysis highlights the importance of the nearest magnetic atomic-layer that manipulates the valley-based topology in Xene. Furthermore, we've also checked the Xene selected as silicene and stanene, with the outcomes shown in Figs. S6-S10. For silicene, eight candidate building-blocks are screened to manifest band-gap-alignment nature within $\pm6\%$ biaxial strains, within which silicene (Si)/NiAs-MnVBi$_2$Te$_5$ and Si/NaCl-MnVBi$_2$Te$_5$ develop $\Gamma$-point based QAHE states with tiny gaps, meanwhile the four Si/Ni$X$Bi$_2$Te$_5$ exhibit valley-based QAHE states [Fig. S6-S8]. For stanene, whereas, it's poor performance existed in stanene (Sn)/$X$Bi$_2$Te$_4$ \cite{li2024multimechanism} is completely inherited into Sn/$XY$Bi$_2$Te$_5$, in which only three band-gap-alignment candidates pass through the $\pm6\%$-strain screening, including Sn/NiAs-MnNiBi$_2$Te$_5$, Sn/NiAs-MnVBi$_2$Te$_5$ and Sn/NaCl-MnVBi$_2$Te$_5$. Among them, only Sn/NiAs-MnVBi$_2$Te$_5$ possess tiny gaps both at $\Gamma$ point and $K$, $K'$ valleys with Chern number as -1 (Fig. S10).

Due to the poor performance of silicene and stanene on both valley gaps and $\Gamma$ gaps, we mainly analyze and discuss the valley-based topological tunability by opting germanene as Xene in the following parts. The valley gap evolving behaviors under various manipulated conditions are mainly discussed in Sec. \ref{iv-B:in-plane} under in-plane magnetism and in Sec. \ref{v:stacking-orders} under out-of-plane magnetism assisted with stacking-order shifts.

\begin{center}
	\begin{figure*}
		\centering
		\includegraphics[width=1\linewidth]{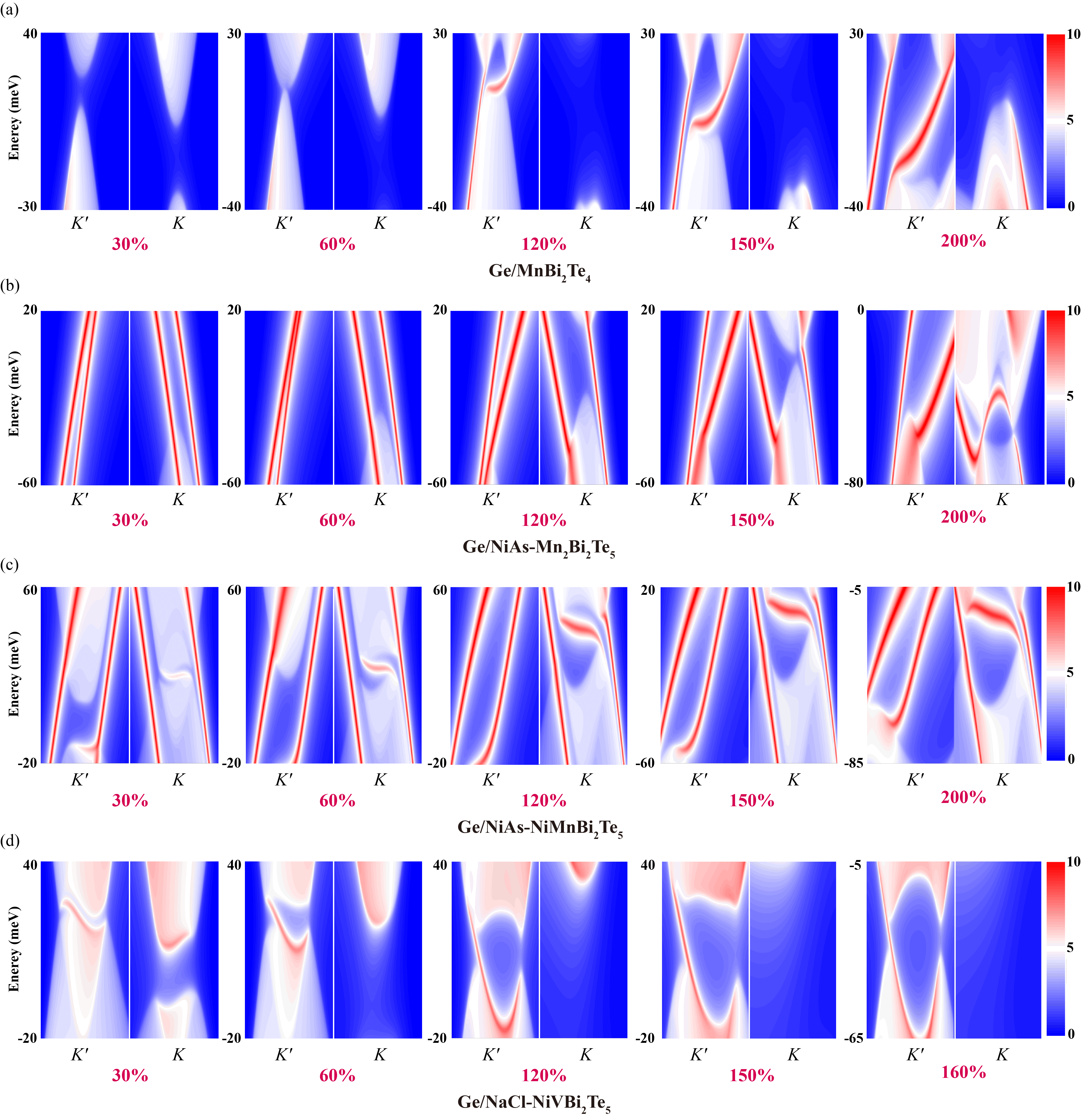}
		\caption{Topological phase transitions of germanene modulated by the SOC strength of its magnetic substrate. (a) Valley-adjacent zoom-in LDOS patterns of Ge/MnBi$_2$Te$_4$ under the SOC strength of the MnBi$_2$Te$_4$ itself varying as 30\%, 60\%, 120\%, 150\% and 200\% that sequentially shown from left to right. (b)-(d) depict similar outcomes, but under the cases of magnetic substrates as NiAs-Mn$_2$Bi$_2$Te$_5$, NiAs-NiMnBi$_2$Te$_5$ and NaCl-NiVBi$_2$Te$_5$. For the case of Ge/NaCl-NiVBi$_2$Te$_5$ under 200\% of the normal substrate-SOC strength, considering the high band-overlapping nature that interferes the valley gaps which is dominated by NaCl-NiVBi$_2$Te$_5$, it is replaced by the condition that under 160\% of the normal substrate-SOC strength.}
		\label{fig:figure3}
	\end{figure*}
\end{center}

 \label{iii:basic_property}
	
\section{Manipulating the Magnetic Substrate}

\subsection{Spin-orbit coupling}	

	SOC plays a crucial role in determining the electronic and topological properties of systems. Typically, enhancing the SOC strength of the substrate enlarges the degree of asymmetry between the two valleys, and then, possibly valley-based QAHE state emerges therein \cite{pan2014valley,pan2015valley}. In this section, we analyze how does varying SOC strength in the magnetic substrate influence the values of valley gap, Berry curvature distributions, and finally the emergence of toplogical phase transition (TPT). The results in the following provide critical insights into the interplay among magnetic elements in substrate composition, inter-vdW hybridization strength and substrate-SOC driven TPTs.
	
	\begin{center}
		\begin{figure*}
			\centering
			\includegraphics[width=1\linewidth]{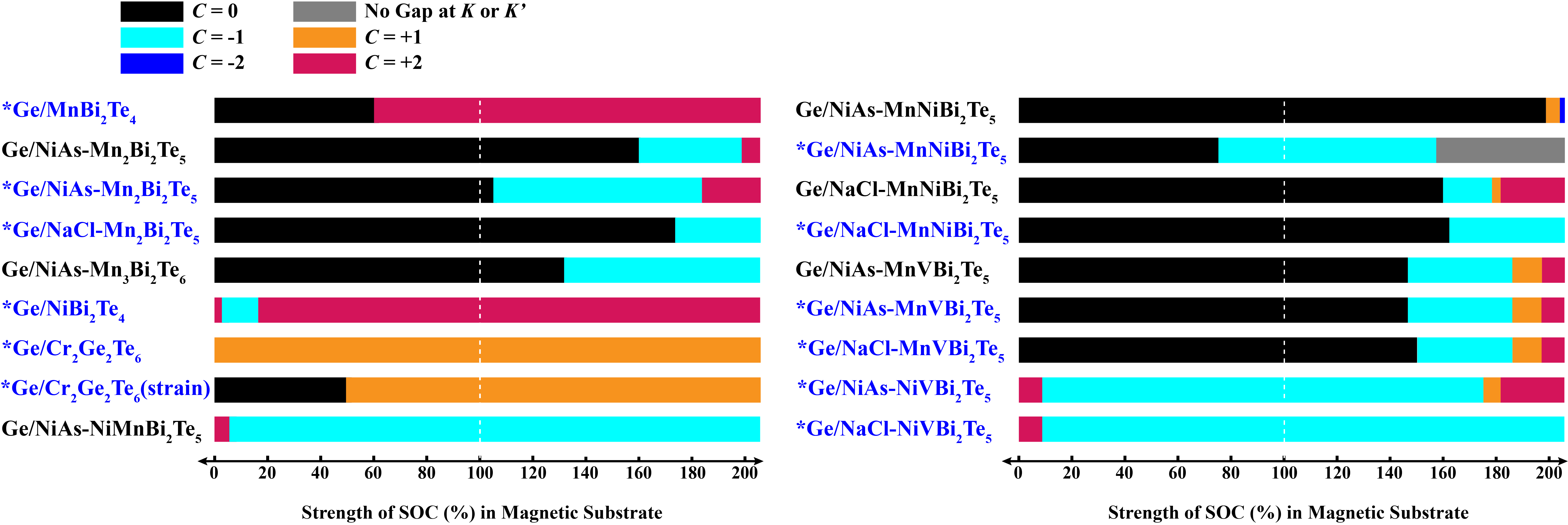}
			\caption{Phase diagram bars of germanene grown on various magnetic substrate that evolve with the SOC strength of the latter. Among these bars, black, blue, cyan, orange, and red colors stand for the Chern number as 0, -2, -1, +1, and +2 accordingly. Gray bar that emerges in the case of FM state Ge/MnNiBi$_2$Te$_5$ means the gap totally vanishes around one of valleys. White vertical dashed line is positioned at the normal strength of substrate-SOC (100\%). In all above 18 selected cases, the magnetic substrate containing AFM coupling is denoted with black, while that under FM couplings is denoted with blue prefixed with a sign “*”.}
			\label{fig:figure4}
		\end{figure*}
	\end{center}

	\begin{center}
		\begin{figure*}
			\centering
			\includegraphics[width=0.95\linewidth]{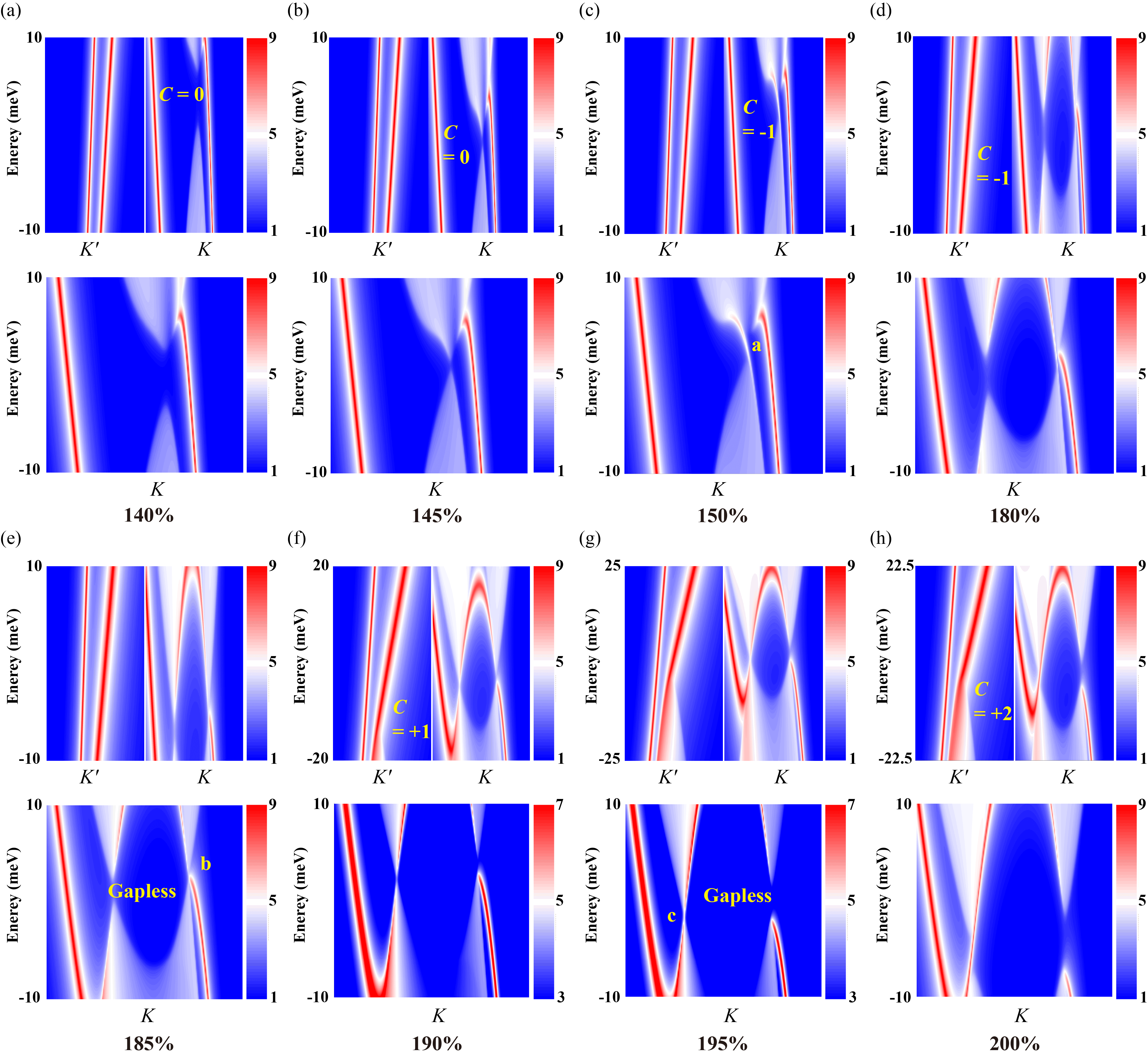}
			\caption{Valley-based topological phase transitions evolve along with the increasing of substrate-SOC strength based on Ge/NiAs-MnVBi$_2$Te$_5$. (a) Under the substrate-SOC strength of 140\%, the higher pattern displays the zoom-in LDOS patterns of Ge/NiAs-MnVBi$_2$Te$_5$ near the $K’$ and $K$ valley gaps, separated with a white vertical cut-off line. The lower pattern exhibits the more detailed zoom-in LDOS pattern around the $K$ valley gap corresponding to the higher pattern. The total Chern number ($C$) is labeled within the no-LDOS zone. Similarly, (b)-(h) show zoom-in LDOS patterns under the substrate-SOC strengths of 145\%, 150\%, 180\%, 185\%, 190\%, 195\% and 200\% respectively.}
			\label{fig:figure5}
		\end{figure*}
	\end{center}

\begin{center}
	\begin{figure*}
		\centering
		\includegraphics[width=1\linewidth]{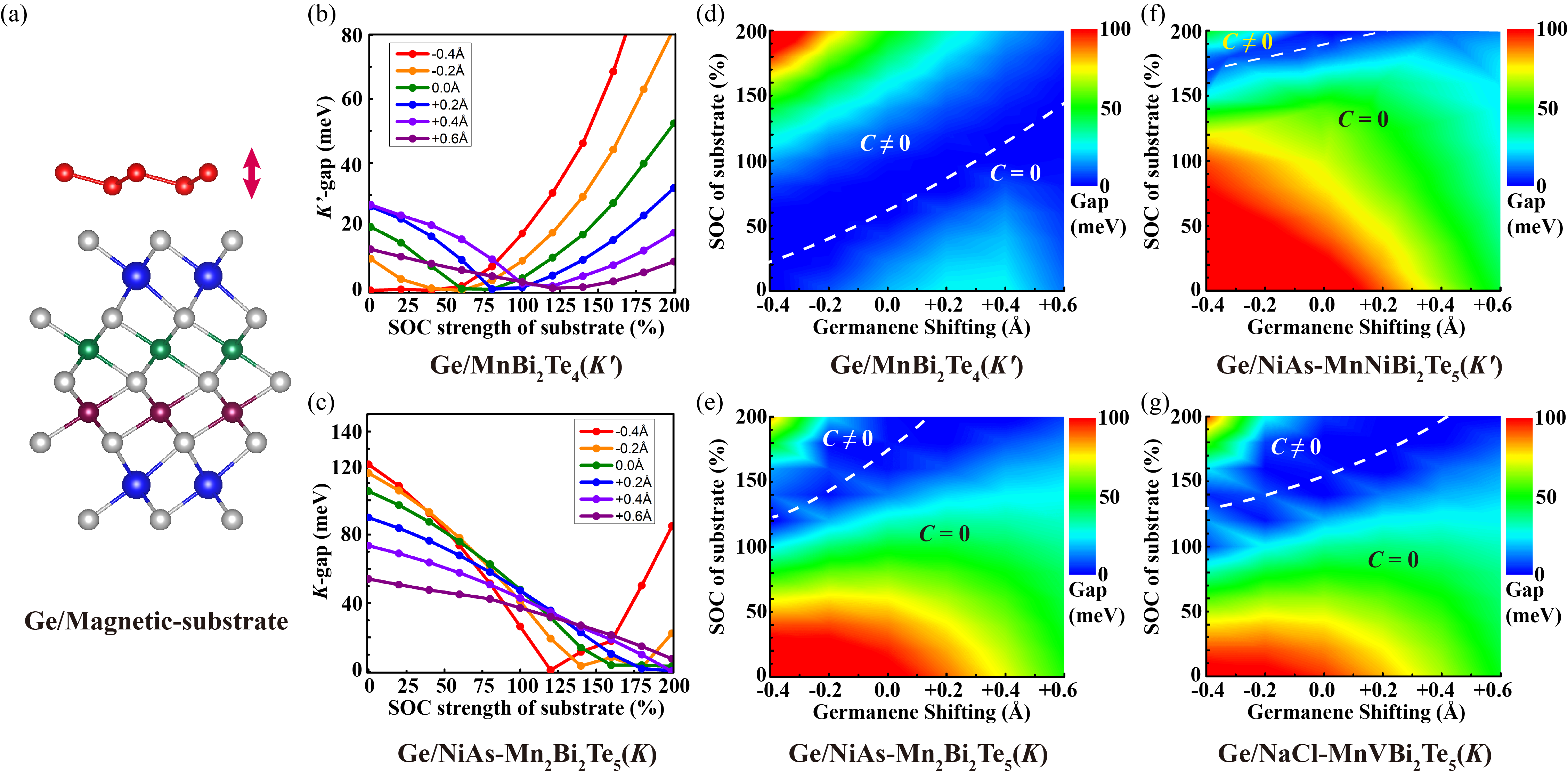}
		\caption{Chern-number and valley-gap phase diagrams by manipulating the inter-vdW spacing distances and the SOC strength of the magnetic substrate. (a) Schematic illustration of out-of-plane shifts between germanene and the magnetic substrate, taking NiAs type $XY$Bi$_2$Te$_5$ as the example. The red two-way arrow in the right of germanene illustrates its shifting direction. (b) In the case of Ge/MnBi$_2$Te$_4$, choosing the $K'$ valley that dedicates the topological phase transition, the gap evolutions along with the substrate-SOC strength are depicted under the inter-vdW spacing distances from -0.4\AA\ to +0.6\AA\ with the colors from red to purple, via setting the balanced inter-vdW spacing distance as 0.0\AA. (c) is similar to that of (b), but under the building block of Ge/NiAs-Mn$_2$Bi$_2$Te$_5$ and the evolution of the $K$ valley gap. (d) Distributions of $K'$ valley gap among the substrate-SOC strength from 0\% to 200\% and the inter-vdW spacing distances from -0.4\AA\ to +0.6\AA\ under the circumstance of Ge/MnBi$_2$Te$_4$. The colors from blue to green then to red are corresponding to the promote of the gap. White dashed bent curve in the mapping pattern positions the topological phase transition between QVHE and QAHE states. (e), (f) and (g) exhibit similar mapping patterns with that of (d), but under the conditions of the $K$ valley gap of Ge/NiAs-Mn$_2$Bi$_2$Te$_5$, the $K'$ valley gap of Ge/NiAs-MnNiBi$_2$Te$_5$ and the $K$ valley gap of Ge/NiAs-MnVBi$_2$Te$_5$ respectively.}
		\label{fig:figure6}
	\end{figure*}
\end{center}

	Figure~\ref{fig:figure3} examines the impact of substrate-SOC on valley-resolved LDOS patterns charactering the toplogical edge states in representative systems. Typically, in Ge/MnBi$_2$Te$_4$ that depicted in Fig. \ref{fig:figure3} (a), although vpQAHE state exists in normal condition \cite{li2024multimechanism,xue2024valley}, decreasing the substrate-SOC strength down to zero totally diminishes the vpQAHE state in germanene, leaving a zero-Chern-number gapped phase. As the substrate-SOC strength is progressively increased, the $K'$ valley gap closes and reopens with Chern number as +2 at about 60\% of normal substrate-SOC, and the valley gap further promotes as the substrate-SOC enhances to even larger degree.
	
	Similar TPT behavior also exists in the Ge/NiAs-Mn$_2$Bi$_2$Te$_5$ system, in which although the normal substrate-SOC strength fails to induce QAHE states in germanene, increasing it up to 200\% recovers the valley-based QAHE state, depicted in Fig. \ref{fig:figure3} (b). As the substrate-SOC increases, the symmetry between the valleys is broken, $K$ valley gap closes and reopens, leading to a Chern-insulating state with $C = +2$. The enhanced SOC-induced asymmetry is attributed to the interplay between the buckled germanene structure and the magnetic coupling in the substrate, which amplifies Rashba spin-orbit effects and then, the asymmetry between two valleys.
	
	If the nearest magnetic atomic-layer is replaced by Ni, the QVHE-QAHE transtion property vanishes along with the substrate-SOC increasing, taken over by metal-QAHE transition with multiple TPT points characterized by different Chern numbers in between. Representatively operated on Ge/NiAs-NiMnBi$_2$Te$_5$, the conductance and valence band around the $K$ valley overlaps setting substrate-SOC strength to 0\%, opening a gap with total Chern number as -1 when the substrate-SOC gradually recovers [Figs. S12(a)-12(c)], then the $K$ valley gap accompanied with the $K'$ gap values promote if the substrate-SOC strength further enlarges within the range below 200\% [Fig. \ref{fig:figure3} (c)]. 
	
	Similar behaviors also performs in Ge/NaCl-NiVBi$_2$Te$_5$ [Fig. \ref{fig:figure3} (d)]. Stemming from the $K$-$K'$ inversion caused by NaCl type stacking structure compared to that of NiAs type, the metal-QAHE transition occurs at $K'$ valley, with the gap-increasing behavior persisting up to 160\% of the substrate-SOC. Above 160\%, the strong overlapping of the band around the $\Gamma$ point hinders the further enlargement of the $K'$ valley gap and even totally destroys it, therefore we only depict the valley-zoom-in LDOS patterns up to 160\% of the substrate-SOC strength in Fig. \ref{fig:figure3} (d).
	
	The substrate-SOC induced TPTs in Xene is not unique to $XY$Bi$_2$Te$_5$-family materials, which can be extended and applied to plenty of other magnetic substrates. In Fig. \ref{fig:figure4} we summarizes the SOC-driven phase transitions across various Ge/magnetic-substrate heterostructures by opting the substrate as $XY$Bi$_2$Te$_5$-family materials, Mn(Ni)Bi$_2$Te$_4$, NiAs-Mn$_3$Bi$_2$Te$_6$ and +2\% biaxial-strain stretched CGT, with the outcomes of TPTs depicted as phase bars. The outcomes reveal that higher SOC strengths generally favor the stabilization of valley-polarized topological states, with variations depending on the substrate's composition and magnetic configuration.

	Obviously, Mn-neared systems experience QVHE-QAHE TPTs as the substrate-SOC enlarges, meanwhile Ni-neared systems generate metal-QAHE TPTs distinctly. For many of them, multiple Chern-number QAHE phases divided by TPTs also occur. For instance, for Ge/NiAs-MnVBi$_2$Te$_5$ that with valley-zoom-in LDOS patterns displayed in Fig. \ref{fig:figure5}, there exists three TPT points, including first TPT point with gap closing-reopening at "a" [Fig. \ref{fig:figure5} (c)] between QVHE and QAHE states ($C = -1$) at 145\% substrate-SOC; second point with gap closing-reopening at "b" [Fig. \ref{fig:figure5} (e)] between QAHE states as "$C = -1$" and "$C = +1$" at 185\%; third point with gap closing-reopening at "c" [Fig. \ref{fig:figure5} (g)] between QAHE states as "$C = +1$" and "$C = +2$" at 195\%. This multiple Chern-insulating phenomenon is originated from that band closing and reopening process of $K$ or $K'$ valleys in these building blocks positions not at the $K$ or $K'$ valley point, but at three points near to the $K$ or $K'$ valley point distributed by $C_{3z}$ rotational symmetry, engendering three closing-reopening points, similar to previous theoretical discussions \cite{pan2015valley}. For some systems, like Ge/MnBi$_2$Te$_4$, the TPT occurs into QAHE with the three gaps closing and reopening simultaneously at around 60\% substrate-SOC strength, immediately pushes the germanene into QAHE state as "$C = +2$". However, for most other building blocks, these three gap-closing and reopening processes don't appear at the same substrate-SOC strength, leaving multiple Chern-insulating phases separated by distinct TPT points. This character indicates us to modulate valley-based Chern numbers by selecting magnetic substrates with various SOC strengths.
	
	Ni-neared building blocks also engender the multiple Chern-insulating phases cut off by different TPT points. Distinctly, the first TPT point appears at very weak substrate-SOC strength. Take Ge/NiAs-NiVBi$_2$Te$_5$ as a representative, shown in Fig. S12, the first TPT point located between 0\% and 20\% substrate-SOC strength puts the system from "$C = +2$" to "$C = -1$". At around 180\%, another TPT occurs, inversing the system back to that $C = +2$.
	
	Notably, replacing the magnetic substrate into monolayer CGT doesn't introduce breakage to the above conclusions. For Ge/CGT under free-relax condition, QAHE state lingers around the substrate-SOC strengths between 0\% and 200\%. Increasing the in-plane lattice constant by 2\% brings TPT point from trivial to QAHE state as "$C = +1$" at 45\% substrate-SOC strength, with the detailed LDOS patterns depicted in Fig. S13. Due to the $\sqrt{3}\times\sqrt{3}$ supercell of germanene that needed to fit the in-plane lattice of CGT, the $K$ and $K'$ valleys folded into $\Gamma$ point, therefore the TPT transition occurs at the $\Gamma$ point [Fig. S13(c)].

The hybridization strength between germanene and the magnetic substrates also plays a critical role in modulating the electronic band structure, valley polarization, and valley-based topological phases, in which tuning the vdW-spacing distance acts as an artificial investigating method to continuously adjusting internal, external Rashba SOC terms and the mass terms imported in germanene. This method serves as an assistance for the convenience of comprehending and designing Xene/magnetic-substrate building blocks by opting appropriate inter-vdW hybridization strengths.

Figure~\ref{fig:figure6} systematically discusses the phase diagram of valley gaps and Chern numbers evolving along with the vdW-spacing distance and substrate-SOC strength. Fig. \ref{fig:figure6}(b) depicts the $K'$ valley gap evolutions of Ge/MnBi$_2$Te$_4$ based on different vdW-spacing distances and substrate-SOC strengths. Among them, 0.0Å stands for the vdW-spacing distance at the equilibrium position, meanwhile other distances are denoted after a subtraction from the distance at the equilibrium position. All of the six vdW-spacing distances manifest non-monotonic evolving behavior of the $K'$ valley gap values, within which the gap closing point moves to higher substrate-SOC strength as the vdW-spacing increases. That is to say, the weaker inter-vdW hybridization strength, the larger substrate-SOC strength is needed to achieve the TPT point. Ge/NiAs-Mn$_2$Bi$_2$Te$_5$ also exhibits very analogous character [Fig. \ref{fig:figure6} (c)], but much larger substrate-SOC strengths are needed. For the vdW-spacing distances larger than 0.0Å, the evolving behavior of $K$ valley gap on which the TPT occurs decays to monotonically decreasing type, indicating that the TPT point of QVHE-QAHE moves above 200\% of substrate-SOC strength.

In order to more intuitively observe the phase diagrams of valley gaps, Chern number evolving along with the factors as vdW spacing distances and substrate-SOC strengths, we depict the valley-gap contour distributions of Ge/MnBi$_2$Te$_4$ at $K'$ valley [Fig. \ref{fig:figure6} (d)], Ge/NiAs-Mn$_2$Bi$_2$Te$_5$ at $K$ valley [Fig. \ref{fig:figure6} (e)], Ge/NiAs-MnNiBi$_2$Te$_5$ at $K'$ valley [Fig. \ref{fig:figure6} (f)] and Ge/NaCl-MnVBi$_2$Te$_5$ at $K$ valley [Fig. \ref{fig:figure6} (g)]. The valley of all candidates above is chosen where the TPT happens, within all of which the magnetic ground state configuration is set.

In each phase diagram, a white dashed curve related to the TPT points exists, dividing the whole contour distributions into "$C = 0$" zone located at the right-bottom part, and "$C \neq 0$" zone located at the left-top part. Blue zone is mainly distributed around the TPT curve. Obviously, the TPT point increases to higher substrate-SOC strength under large vdW-spacing distance regimes, in which in Ge/NiAs-Mn$_2$Bi$_2$Te$_5$, Ge/NiAs-MnNiBi$_2$Te$_5$ and Ge/NaCl-MnVBi$_2$Te$_5$ the TPT point moves above 200\% substrate-SOC as the vdW-spacing rise above +0.4Å. Noticeably, large valley gaps with nonzero Chern numberg mainly gather at the red-colored, left-top part, suggesting that large gapped valley-based QAHE state's appearance favors strong inter-vdW hybridizations and large substrate-SOC strength.

The phase diagrams displayed among Figure~\ref{fig:figure6} summarize the correlation between vdW spacing distances, substrate-SOC strengths and topological phase transitions across various heterostructures. Besides, we also draw the phase diagrams of Ni-neared system, in which $K$ valley gap of Ge/NiAs-NiMnBi$_2$Te$_5$ and Ge/NiAs-NiVBi$_2$Te$_5$ are included in Fig. S14. The distribution of valley gaps grow more complicated, within which the large gapped QAHE state still gathers at the left-top part of each phase diagram.

In conclusion, substrate-SOC strength acts as a vital rule determining valley-based topological properties in Xene/magnetic-substrate heterostructures. By experimentally modulating substrate-SOC strength and the inter-vdW hybridization strength, i. e., by selecting magnetic substrates that containing various SOC strength and inter-vdW coupling strength with Xene, it is possible to precisely control valley-based QAHE states (in most conditions, vpQAHE states), achieve large valley gaps, and explore new valley-based topological phases.

 \label{iv-A:substrate-soc}
	
\subsection{In-plane magnetic moment orientation}

From a theoretical perspective, the ability of in-plane magnetic moments to induce QAHE depends critically on their direction relative to the in-plane mirror symmetry of crystal lattice in the latter of which the band closing-reopening process rises \cite{liu2013plane,ren2016quantum,liu2018intrinsic}. Specific in-plane orientations effectively break all the three in-plane mirror symmetries that distributed according to $C_{3z}$ rotational symmetry, enabling Berry curvature asymmetry and valley-based QAHE, while others restore certain mirror symmetry, suppressing valley-based QAHE state and leaving with gapless nature in one of the valley. This sensitivity underscores the importance of controlling magnetic moment orientation in experimental studies of 2D topological materials.

		\begin{center}
		\begin{figure*}
			\centering
			\includegraphics[width=0.81\linewidth]{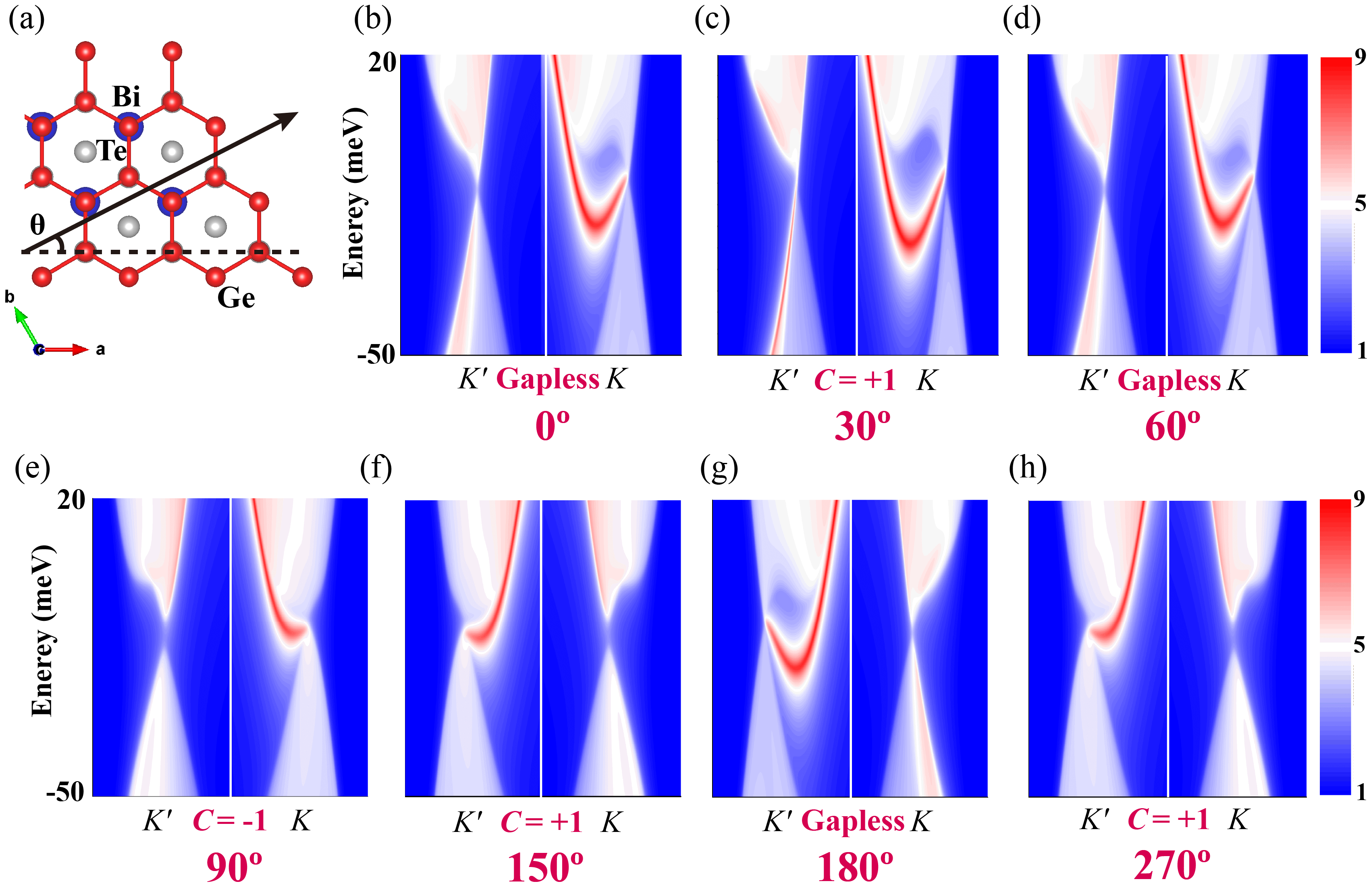}
			\caption{Valley-gap and Chern-number evolutions along with the in-plane magnetic moment orientations based on representative Ge/NaCl-NiVBi$_2$Te$_5$, depicted detailly with $K$($K’$) valleys neighbored zoom-in LDOS patterns by choosing seven certain magnetic-moment directions as example. (a) Top view of Ge/NaCl-NiVBi$_2$Te$_5$ in which the red, blue and light-gray balls stand for Ge, Bi and Te atoms. Horizontal black dashed line is parallel to the axis \textit{\textbf{a}}, meanwhile the black arrow illustrates the magnetic-moment direction. The included angle $\theta$ defines the angle of magnetic-moment direction that employed in (b)-(h). From (b) to (h) the angle $\theta$ is chosen as $0^{\circ}$, $30^{\circ}$, $60^{\circ}$, $90^{\circ}$, $150^{\circ}$, $180^{\circ}$ and $270^{\circ}$ respectively. Non-zero Chern numbers or gapless phase is denoted with red below each pattern, while the zero-Chern-number gapped phase is denoted with black. All the LDOS patterns are extracted from the [100] edge of the system.}
			\label{fig:figure7}
		\end{figure*}
	\end{center}

	\begin{center}
		\begin{figure*}
			\centering
			\includegraphics[width=1\linewidth]{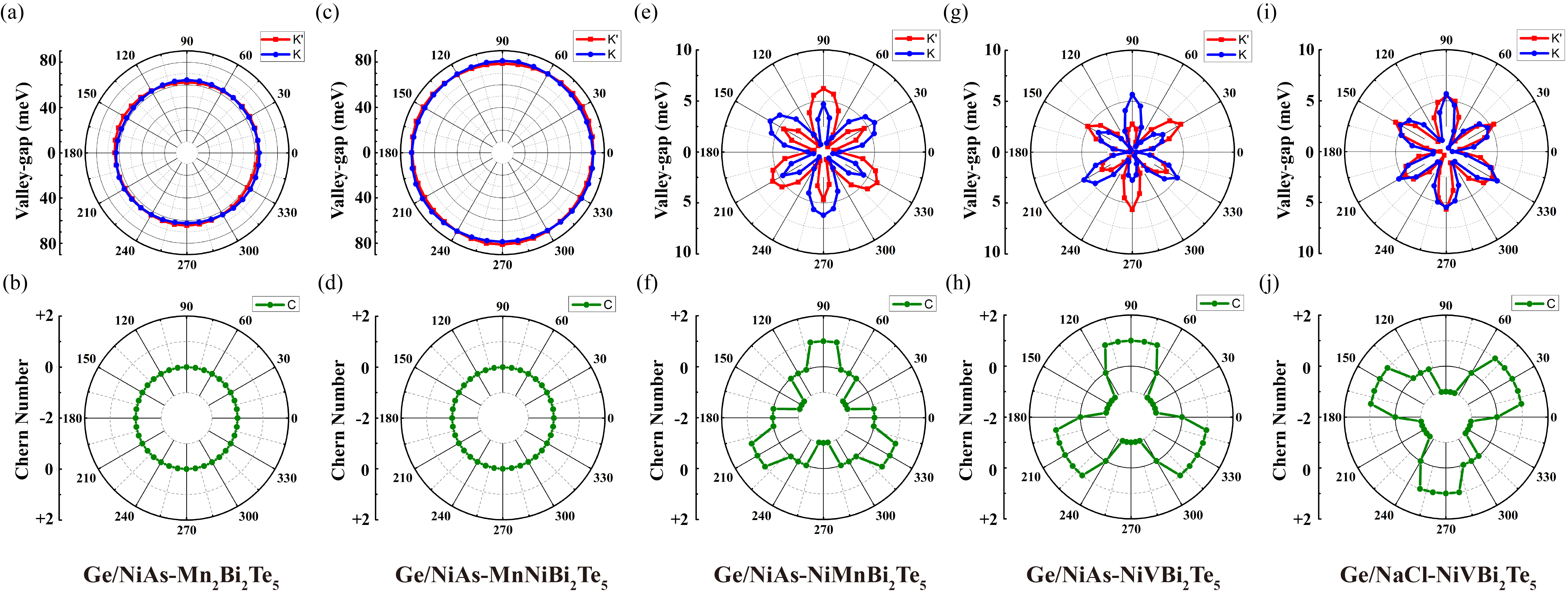}
			\caption{Valley-gap and Chern-number evolutions along with the in-plane magnetic moment orientations. (a) In-plane magnetic moment orientation induced $K'$ valley (red curve) and $K$ valley (blue curve) gap evolutions by opting Ge/NiAs-Mn$_2$Bi$_2$Te$_5$. Its Chern number evolution is also depicted in (b). Homoplastically, (c) and (d), (e) and (f), (g) and (h), (i) and (j) depict the valley-gap and the Chern-number evolutions of Ge/NiAs-MnNiBi$_2$Te$_5$, Ge/NiAs-NiMnBi$_2$Te$_5$, Ge/NiAs-NiVBi$_2$Te$_5$ and Ge/NaCl-NiVBi$_2$Te$_5$ accordingly.}
			\label{fig:figure8}
		\end{figure*}
	\end{center}
	
Here, the orientation of in-plane magnetic moments introduces rich dynamics in Ge/mangetic-substrate heterostructures, particularly in modulating valley gaps, Berry curvature, and Chern numbers ($C$). First of all, by taking Ge/NaCl-NiVBi$_2$Te$_5$ as a representative, displayed in Fig. \ref{fig:figure7}, we select seven in-plane magnetic orientation angles: $0^{\circ}$, $30^{\circ}$, $60^{\circ}$, $90^{\circ}$, $150^{\circ}$, $180^{\circ}$ and $270^{\circ}$ coming up with valley-around zoom-in LDOS patterns, depicted sequentially with Figs. \ref{fig:figure7} (b)-\ref{fig:figure7}(h). The orientation angle of the magnetic moment is defined as $\theta$ with the intersection angle from the $\textit{\textbf{a}}$ axis, illustrated in Fig. \ref{fig:figure7} (a). Gapless state emerges at $0^{\circ}$, $60^{\circ}$ and $180^{\circ}$ shown in Fig. \ref{fig:figure7}, with one of the three mirror symmetry protected. The Chern number develops as +1 at $30^{\circ}$, -1 at $90^{\circ}$, +1 at $150^{\circ}$ respectively, confirming the in-plane magnetic orientation manipulated Chern number tunable and chirality switchable nature.

Fig. \ref{fig:figure8} provides a comprehensive overview of the $K$ and $K'$ valley-gap evolution and Chern number transitions under varying in-plane magnetic orientations under normal substrate-SOC strength, focusing on the five opted systems involving Ge/NiAs-Mn$_2$Bi$_2$Te$_5$, Ge/NiAs-MnNiBi$_2$Te$_5$, Ge/NiMnBi$_2$Te$_5$, Ge/NiAs-NiVBi$_2$Te$_5$ and Ge/NaCl-NiVBi$_2$Te$_5$. For Mn-neared systems, derived from representative Ge/NiAs-Mn$_2$Bi$_2$Te$_5$ and Ge/NiAs-MnNiBi$_2$Te$_5$, the $K$ and $K'$ valley gaps share almost the same value with the anisotropy almost absent [Figs. \ref{fig:figure8} (a) and \ref{fig:figure8} (c)], and with the Chern number totally fixed to zero [Figs. \ref{fig:figure8} (b) and \ref{fig:figure8} (d)].

Conversely, the Ni-neared systems behave more complicated and interested due to their nonzero Chern-insulating character [Figs. \ref{fig:figure8}(e)-\ref{fig:figure8}(j)], with the valley gaps closing at $0^{\circ}$, $60^{\circ}$, $120^{\circ}$, $180^{\circ}$, $240^{\circ}$ and $300^{\circ}$, in the angle of which one of the in-plane mirror symmetry maintains. These valley gap closing points are intuitively related to TPT points with Chern number alterations. The valley gaps of $K$ and $K'$ valley obey the in-plane inversion-symmetry rule, explaining more concretely, the gap value of $K$ ($K'$) valley equals to that of $K'$ ($K$) valley after the in-plane magnetic moment orientated by $180^{\circ}$. The Chern number also abides the above rule, in which the Chern number falls to zero around the above six mirror-symmetry related angles, and manifests itself as +1 or -1 alternately at other angles of in-plane magnetic orientation. The above novel behaviors remains among the selected representative cases in the main text as Ge/NiAs-NiMnBi$_2$Te$_5$ [Figs. \ref{fig:figure8}(e) and \ref{fig:figure8}(f)], Ge/NiAs-NiVBi$_2$Te$_5$ [Figs. \ref{fig:figure8}(g) and \ref{fig:figure8}(h)] and Ge/NaCl-NiVBi$_2$Te$_5$ [Figs. \ref{fig:figure8}(i) and \ref{fig:figure8}(j)].

\begin{center}
	\begin{figure}
		\centering
		\includegraphics[width=1\linewidth]{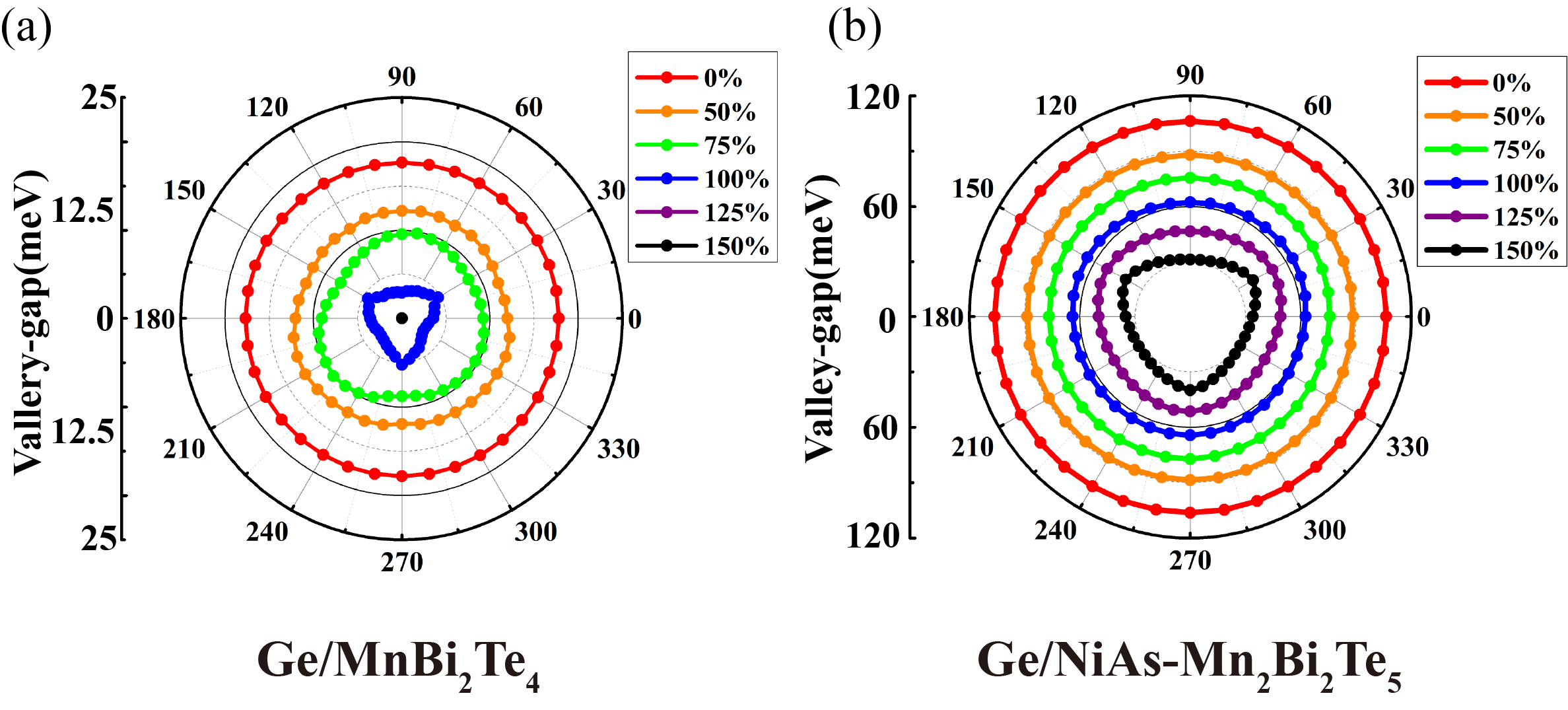}
		\caption{Valley-gap evolutions manipulated by in-plane magnetic orientations under various substrate-SOC strengths under (a) Ge/MnBi$_2$Te$_4$ and (b) Ge/Mn$_2$Bi$_2$Te$_5$. Red, orange, green, blue, purple and black curves are derived under the substrate-SOC strengths of 0\%, 50\%, 75\%, 100\%, 125\% and 150\%. Both of the valley gaps are extracted from the $K’$ valley. }
		\label{fig:figure9}
	\end{figure}
\end{center}

\begin{center}
	\begin{figure*}
		\centering
		\includegraphics[width=1\linewidth]{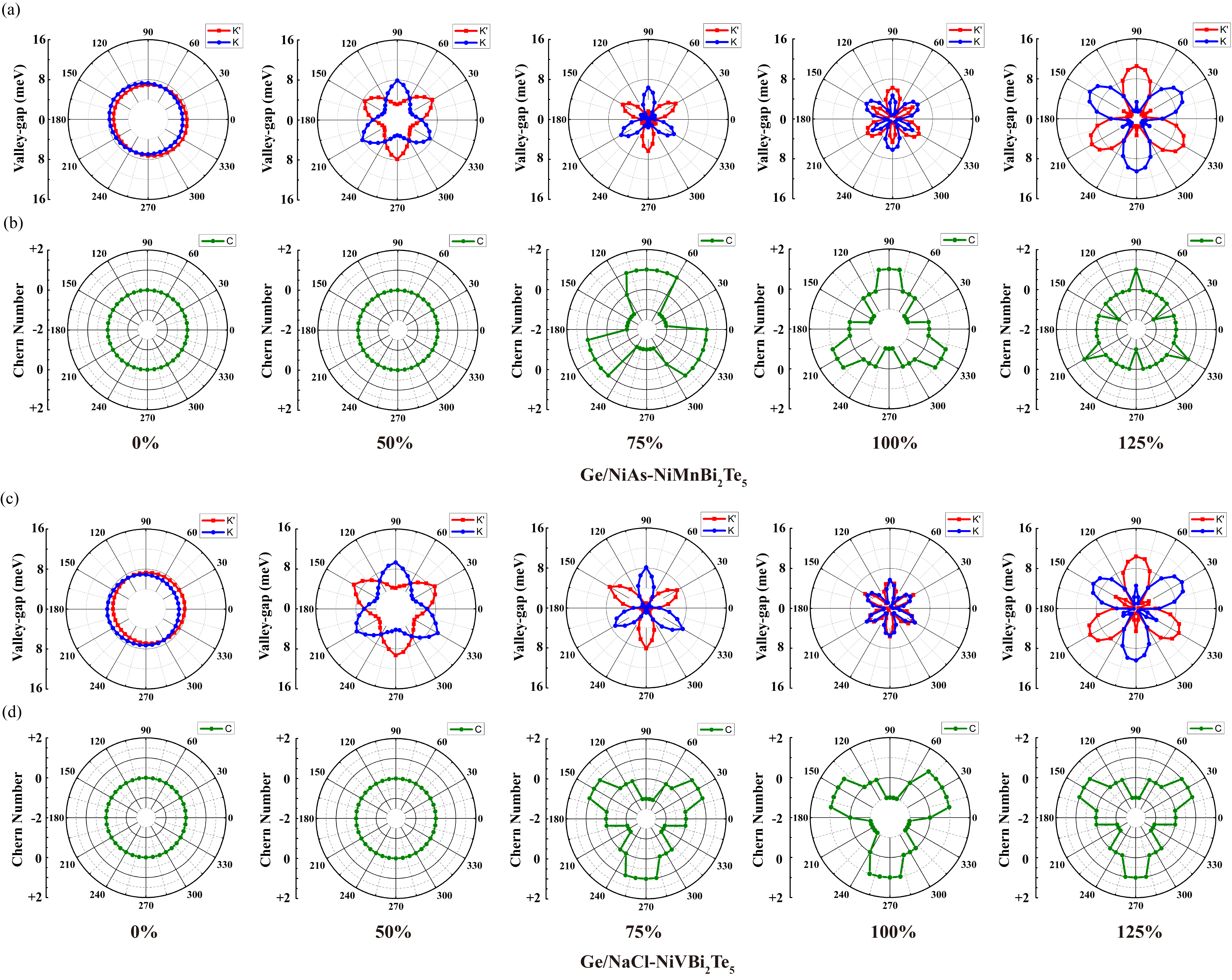}
		\caption{Valley-gap and Chern-number evolutions along with the in-plane magnetic moment orientations under various substrate-SOC strengths under cases of (a)-(b) Ge/NiMnBi$_2$Te$_5$ and (c)-(d) Ge/NaCl-NiVBi$_2$Te$_5$. In all the transversely aligned patterns, from left to right the SOC strength increases as 0\%, 50\%, 75\%, 100\% and 125\%.}
		\label{fig:figure10}
	\end{figure*}
\end{center}
	
Notably, the substrate-SOC strength also crucially determines the above in-plane magnetic orientation manipulated valley-based topology. For Mn-neared systems discussed in Fig. \ref{fig:figure8}, promoting the substrate-SOC strength reduces the valley gap down to metallic phase under in-plane magnetism, no Chern-number TPT occurs, therefore we only depict the $K'$ valley gap evolutions driven by in-plane magnetic orientations under 0\%, 50\%, 75\%, 100\%, 125\% and 150\% substrate-SOC strengths with red, orange, green, blue, purple and black curves respectively in Fig. \ref{fig:figure9}. For both Ge/MnBi$_2$Te$_4$ and Ge/NiAs-Mn$_2$Bi$_2$Te$_5$, the enhancement of substrate-SOC promotes the in-plane anisotropy of the valley gaps, but shrinks and even closes them at strong substrates-SOC regime. Hence, it's not an excellent designing routine to acquire in-plane magnetism and valley-based QAHE states.

In contrast, Ni-neared systems habour more plentiful outcomes of valley-based topological tunability. As usual, we also adopt two typical Ni-neared candidates: Ge/NiAs-NiMnBi$_2$Te$_5$ and Ge/NaCl-NiVBi$_2$Te$_5$, with in-plane magnetism based valley-gap and Chern number evolutions displayed in Fig. \ref{fig:figure10}. For both conditions, we also select 0\%, 50\%, 75\%, 100\% and 125\% substrate-SOC strengths. 

The two candidate systems bear very similar properties, therefore we handle and discuss them together. The absence of substrate-SOC (0\%) destroyes the valley-gap anisotropy, forms circle-like pattern [Figs. \ref{fig:figure10} (a) and \ref{fig:figure10}(c)] with totally zero Chern number [Figs. \ref{fig:figure10} (b) and \ref{fig:figure10}(d)]. If half of the substrate-SOC strength recovers (50\%), in-plane anisotropy clearly introduced into valley-gap evolutions obeying three-fold rotational symmetry [the second patterns of Figs. \ref{fig:figure10} (a) and \ref{fig:figure10}(c)], the Chern number still fixes to zero without any TPTs. 75\% of substrate-SOC successfully imports the QAHE state (alternated by Chern numbers as +1 and -1) into germanene at certain magnetic orientations, simultaneously appears with valley-gap closing points distributed with six-fold symmetry. Nevertheless, the gaps oscillate within small regime (the maximum up to 6meV) and also rely on three-fold rotational symmetry (the third pattern in each sub-figure of Fig. \ref{fig:figure10}). Moreover, further enhancing the substrate-SOC to 125\% enlarges the values of valley-gap in some magnetic orientations, with the maximum up to about 10meV, but shrinks the non-zero Chern number angle ranges [the right edge pattern in each sub-figure of Fig. \ref{fig:figure10}]. 150\% of substrate-SOC strength ruins the Chern-insulating characters at all the angles. Consequently, for Ni-neared systems, controlling the moderate substrate-SOC strength is vital for obtaining Chern number tunable and chirality switchable characters based on orientations of in-plane magnetism.

Detailed valley-neared zoom-in LDOS patterns of Ge/NiAs-NiMnBi$_2$Te$_5$ and Ge/NaCl-NiVBi$_2$Te$_5$ under the above five substrate-SOC strengths are exhibited sequentially in Figs. S15 and S16 respectively, both under angles of $10^{\circ}$, $30^{\circ}$, $50^{\circ}$, $70^{\circ}$, $90^{\circ}$ and $110^{\circ}$. These LDOS patterns intuitively exhibit the Chern insulating behavior that varies alternately with each other between the two valleys under the substrate-SOC range from 75\% to 125\%.

In summary, we've comprehensively analyzed the in-plane magnetic orientation and substrate-SOC manipulated valley-based QAHE states by adopting five representative Ge/$XY$Bi$_2$Te$_5$ building-blocks. Entirely, enhancing the substrate-SOC strength imports the in-plane-magnetism induced anisotropy into the valley-gap distribution of germanene, decreasing the gaps of both valleys, but may not induces the valley-based QAHE state. For Mn-neared cases, metallic phase instead of QAHE state emerges under strong substrate-SOC condtions. Moderate substrate-SOC strength is necessary for importing Chern number alternate and chirality switchable characters into Ni-neared germanene based on in-plane magnetic orientations, with valley gaps closing and reopening at in-plane mirror-symmetry protected angles. Too strong or too weak substrate-SOC destroys the nonzero Chern insulating nature, which is crucial for theoretical and experimental magnetic-substrate selections on further surveys and applications. This section firstly provides an intuitive guiding principle for designing and manipulating in-plane magnetism induced valley-based QAHE states in Xene via selecting and manipulating appropriate substrates with their magnetic properties. 

 \label{iv-B:in-plane}
 
  \label{iv:substrate-manipulation}

\section{Stacking-order Shifts}
	
\begin{center}
	\begin{figure*}
		\centering
		\includegraphics[width=1\linewidth]{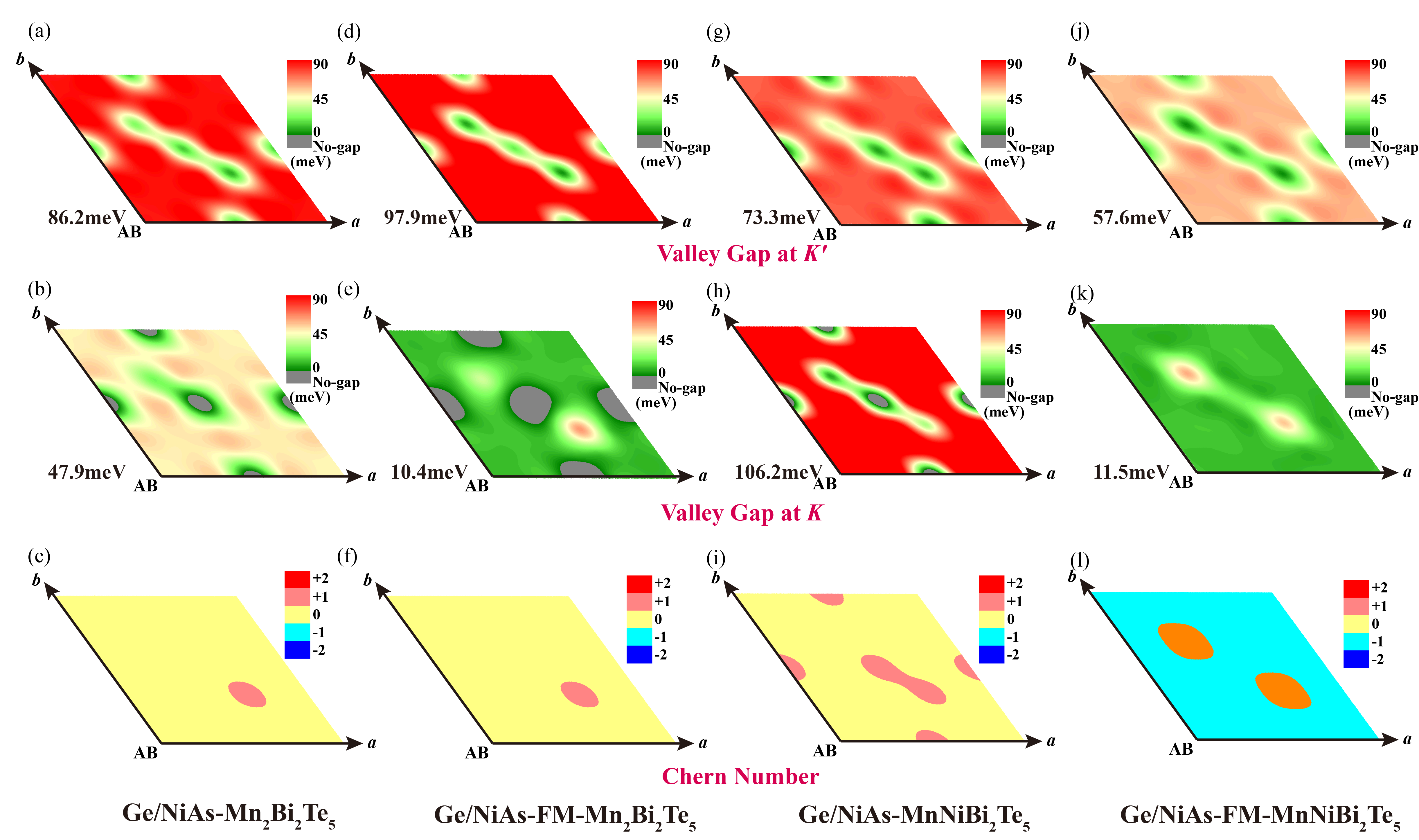}
		\caption{Valley gaps and the Chern numbers manipulated by stacking-order shifts under the cases of (a)-(c) Ge/NiAs-Mn$_2$Bi$_2$Te$_5$, (d)-(f) FM-state Ge/NiAs-Mn$_2$Bi$_2$Te$_5$, (g)-(i) Ge/NiAs-MnNiBi$_2$Te$_5$, and (j)-(l) FM-state Ge/NiAs-MnNiBi$_2$Te$_5$. In valley gap distributions, from green to yellow then to red colors the gap promotes, within which the gray zone means no gap exists here. In Chern number distributions, blue, cyan, yellow, orange and red colors expound the Chern numbers: -2, -1, 0, +1, +2 respectively. The gap values of the normal stacking orders are denoted in the left side to the point "$AB$" of valley-gap distribution patterns.}
		\label{fig:figure11}
	\end{figure*}
\end{center}

\begin{center}
	\begin{figure*}
		\centering
		\includegraphics[width=1\linewidth]{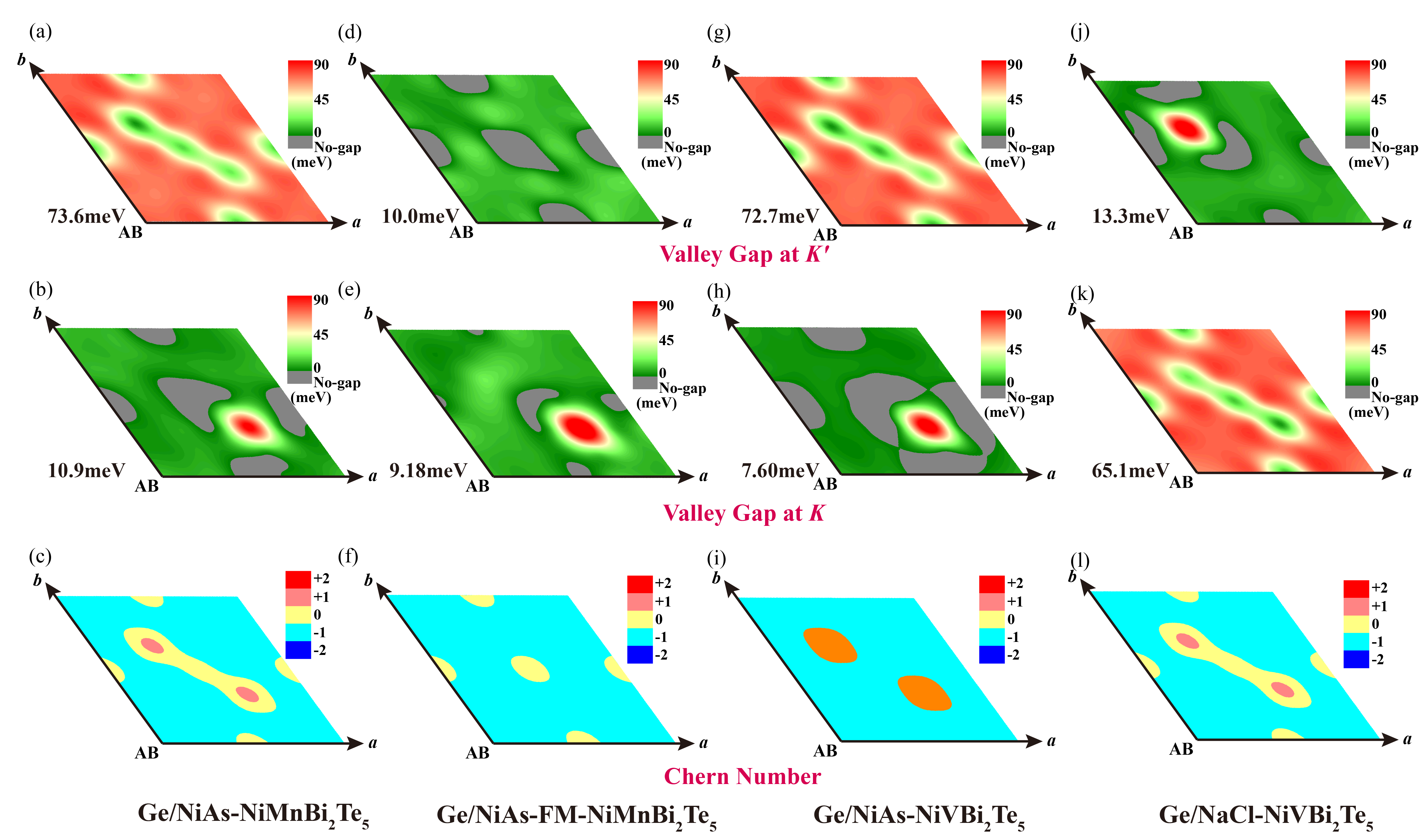}
		\caption{Valley gaps and the Chern numbers manipulated by stacking-order shifts under the cases of (a)-(c) Ge/NiAs-NiMnBi$_2$Te$_5$, (d)-(f) FM-state Ge/NiAs-NiMnBi$_2$Te$_5$, (g)-(i) Ge/NiAs-NiVBi$_2$Te$_5$, and (j)-(l) Ge/NaCl-NiVBi$_2$Te$_5$.}
		\label{fig:figure12}
	\end{figure*}
\end{center}

In this section, we foucs on the valley gaps of Xene/magnetic-substrate mainly discussed in this work that perform as an fundamental role determining the realizing temperatures of their valley-based topological features, analyzed and optimized by the stacking-order-shift method. The stacking-order shift of Xene relative to the magnetic substrates provides another essential degree of freedom for tuning the valley gaps and valley-based topological phases of the heterostructures \cite{li2024multimechanism}. Unlike the structural differentia between NiAs type and NaCl type configurations in the substrate itself, the stacking-order shift refers to the in-plane relative displacement (shift, not twist) between germanene and the substrate. This displacement breaks in-plane symmetry, alters the electronic hybridization strength between Xene and the magnetic substrate, and significantly affects the valley-based Berry curvature distribution and the topological phase transitions imported into the Xene. Figures~\ref{fig:figure11} and \ref{fig:figure12} analyze the impact of stacking-order shifts on valley gaps and the chirality, the values of Chern numbers across various representative systems based upon out-of-plane magnetism within the substrates.

Figure~\ref{fig:figure11} presents the gap distributions at the $K$ and $K'$ valleys accompanied with Chern number for Mn-neared germanene-based heterostructures, involving Ge/NiAs-Mn$_2$Bi$_2$Te$_5$, FM state Ge/NiAs-Mn$_2$Bi$_2$Te$_5$, Ge/NiAs-MnNiBi$_2$Te$_5$, and FM state Ge/NiAs-MnNiBi$_2$Te$_5$. In these systems that harbours QVHE state at normal stacking conditions, the gap at $K'$ valley all arrive at large regime (all above 50meV), and only shrinks to small regime (around 10meV) at $K$ valley when the interatomic-layer interactions reverse to FM. Stacking-order shift reduces the $K'$ valley gap to about 10meV along the middle zones around the [$1\overline{1}0$] positions, with the $K$ valley gap down to the metallic properties around these zones. In most other areas, the large gapped nature persists. These behaviors perform likely with that of Ge/MnBi$_2$Te$_4$ \cite{li2024multimechanism}.
	
Comparing the two magnetic configurations (FM and AFM) of Ge/NiAs-Mn$_2$Bi$_2$Te$_5$, exhibited in Figs. \ref{fig:figure11} (a)-\ref{fig:figure11}(c) related to AFM state and  Figs. \ref{fig:figure11} (d)-\ref{fig:figure11}(f) related to FM state, the magnetic-moment inversion of the second nearest magnetic layer leaves no influence on the Chern number distributions, both of which generate QVHE state in most zone but QAHE state as "$C = +1$" around the (2/3, 1/3) shifting point. However, its $K$ valley gap suffers obviously to the second nearest magnetic moment, effecting through the short distance of NiAs type based Mn-Mn direct interaction. Hence, the QVHE-QAHE TPT point happens at weaker substrate-SOC strength under FM state (about 110\%) than that under AFM state (about 160\%), clearly seen from the second and third bar in left panel of Fig. \ref{fig:figure4}.

For the case of Ge/NiAs-MnNiBi$_2$Te$_5$, the magnetic-moment inversion of the second nearest magntic Ni layer not only reforms the $K$ and $K'$ valley gap distributions, but also transform the QVHE into QAHE state in most zones with the Chern number as -1 [Fig. \ref{fig:figure11}(l)]. The short Ni-Mn direct interaction distance fails to decay the strong Ni-based effects, therefore leads to entirely TPTs in most stacking-order zones.

We also make trials on Ge/MnVBi$_2$Te$_5$ under NiAs type, FM coupling of NiAs type and NaCl type conditions in Fig. S17. The second nearest V atom contributes very weak magnetic interactions to germanene, hence causes almost little influence on even the valley gaps by comparing the cases of Ge/NiAs-MnVBi$_2$Te$_5$ [Figs. S17(a)-S17(c)] to FM-state Ge/NiAs-MnVBi$_2$Te$_5$ [Figs. S17(d)-S17(f)]. The Chern number distributions under these two conditions are same to those Ge/NiAs-Mn$_2$Bi$_2$Te$_5$ and AFM state Ge/NiAs-MnNiBi$_2$Te$_5$. Changing the NiAs type structure of MnVBi$_2$Te$_5$ into NaCl type brings no reformation on valley-gap and Chern number distributions, but the spatial reversion between $K'$ and $K$ valleys, and spatial distributed inversion of the Chern number [Figs. S17(g)-S17(i)].

Figure~\ref{fig:figure12} extends the stacking-order analysis to Ni-neared building-blocks, including Ge/NiAs-NiMnBi$_2$Te$_5$, Ge/NiAs-FM-NiMnBi$_2$Te$_5$, Ge/NiAs-NiVBi$_2$Te$_5$ and Ge/NaCl-NiVBi$_2$Te$_5$, with more plentiful manipulated valley-based topology arises. We also take examples of Ge/NiAs-NiMnBi$_2$Te$_5$ firstly in Figs. \ref{fig:figure12}(a)-\ref{fig:figure12}(f), in which the second-nearest Mn layer also acts weakly on the Chern number distributions, and strongly on the $K'$ valley gaps. For the AFM state Ge/NiAs-NiMnBi$_2$Te$_5$, the Chern number can be step likely modulated as -1, 0 and +1 [Fig. \ref{fig:figure12}(c)].
	
Replacing the magnetic substrate by NiVBi$_2$Te$_5$ provides alike outcomes, with the smaller gapped valley distributes around 0 to 15meV in most area, but a small large gapped zone up to 90meV at the (2/3, 1/3) point for NiAs type and (1/3, 2/3) point for NaCl type structures respectively. Only Ge/NaCl-NiVBi$_2$Te$_5$ has the ability of step likely modulating Chern number as -1, 0, +1 [Fig. \ref{fig:figure12}(k)] meanwhile in Ge/NiAs-NiVBi$_2$Te$_5$ the zero Chern number phase totally vanishes [Fig. \ref{fig:figure12}(h)].

\begin{center}
	\begin{figure*}
		\centering
		\includegraphics[width=0.8\linewidth]{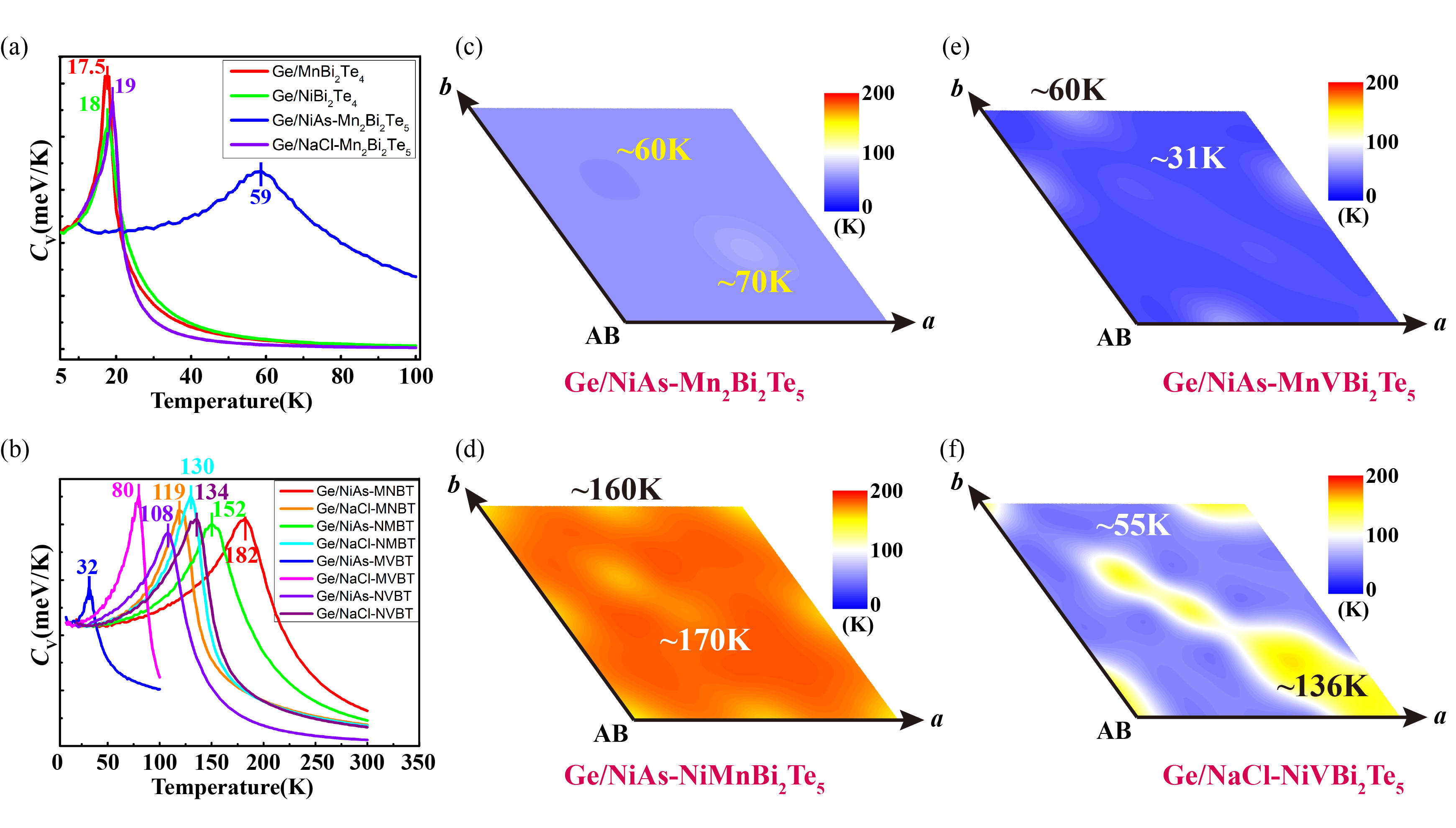}
		\caption{Magnetic critical temperatures with their stacking-order modulations under various representative building blocks of Ge/magnetic-substrate. (a) and (b) unfold $C_{\rm V}$-$T$ curves of Ge/$X$Bi$_2$Te$_4$ ($X$ = Mn, Ni) and Ge/$XY$Bi$_2$Te$_5$ depicted by multiple distinct colors. Digits marked at the peak of each curve sharing the same color give critical temperature values of each case. Stacking-order shift induced $T_{\rm C}$ modulations under the cases of (c) Ge/NiAs-Mn$_2$Bi$_2$Te$_5$, (d) Ge/NiAs-NiMnBi$_2$Te$_5$, (e) Ge/NiAs-MnVBi$_2$Te$_5$ and (f) Ge/NaCl-NiVBi$_2$Te$_5$ revealed with mapping distributions. Colors from blue to white, yellow then to red, the critical temperature increases. For all of these four cases, the distributions of critical temperature mainly fall into two zones, both of which are denoted with the mean values.}
		\label{fig:figure13}
	\end{figure*}
\end{center}

Akin to the properties of Ge/MnBi$_2$Te$_4$ \cite{li2024multimechanism}, stacking-order shift performs as experimentally executable way to manipulate and optimize both valley gaps and Chern numbers on germanene grown on other magnetic substrates. Due to the valley asymmetry imported by the substrate, in most conditions the gap of one of the valleys inevitably shrinks to small regime (the smallest of selected system down to 7.6meV), within which reforming the stacking order can slightly optimize it to higher values. This gap value can be obviously reformed by changing the magnetic configurations of the substrate, and high enough to support high-temperature valley-based topological features to about or above 90K assuming that its magnetic critical temperature falls above this value, majorly discussed in the Sec. \ref{vi:tc}. Besides, the chirality and values of Chern number can also be manipulated by stacking-order shifts, among which Ge/NiAs-NiMnBi$_2$Te$_5$ and Ge/NaCl-NiVBi$_2$Te$_5$ harbours the ability of modulating Chern numbers as -1, 0, +1.

 \label{v:stacking-orders}
	
\section{Magnetic Critical Temperature}

$XY$Bi$_2$Te$_5$-family materials possess appealing potential for high magnetic critical temperature \cite{tang2023intrinsic} benefiting from their biatomic-layer magnetic structures with relatively strong interatomic magnetic coupling, which greatly supresses the 2D magnetic fluctuations. Hence, we can expect the Xene/$XY$Bi$_2$Te$_5$ that supports high temperature valley-based topology.

Figure~\ref{fig:figure13} illustrates the magnetic critical temperatures (\(T_C\)) and their stacking-order manipulations for various representative Ge/magnetic-substrate heterostructures. Figs. \ref{fig:figure13} (a) and (b) present the heat capacity (\(C_V\)) versus temperature (\(T\)) curves for germanene-based heterostructures. In each curve, values that denoted near the peak highlights the critical temperature. Notably, in Fig. \ref{fig:figure13} (a) we draw the $C_V - T$ curves of Ge/MnBi$_2$Te$_4$, Ge/NiBi$_2$Te$_4$, Ge/NiAs-Mn$_2$Bi$_2$Te$_5$ and Ge/NaCl-Mn$_2$Bi$_2$Te$_5$ colored with red, green, blue and purple respectively, with Curie temperatures being 17.5K, 18K, 59K and 19K accrodingly. Fig. \ref{fig:figure13} (b) reveals $C_V - T$ curves under other Ge/$XY$Bi$_2$Te$_5$ materials mentioned in this work. Not surprisingly, the influence of $T_{\rm C}$ from germanene layer is almost ignorable compared to free-standing $XY$Bi$_2$Te$_5$, the highest $T_{\rm C}$ of four building blocks arriving as 182K for Ge/NiAs-MnNiBi$_2$Te$_5$, 152K for Ge/NiAs-NiMnBi$_2$Te$_5$, 134K for Ge/NaCl-NiVBi$_2$Te$_5$ and 130K for Ge/NaCl-NiMnBi$_2$Te$_5$. For more clarity we list all the results of $T_{\rm C}$ values in Table \ref{tab2:Tc} mentioned in Figs. \ref{fig:figure13}(a) and \ref{fig:figure13}(b).

\begin{table}
	\caption{\label{tab2:Tc} The $T_{\rm C}$ values of Ge/$XY$Bi$_2$Te$_5$ that discussed in this work, compared with that of Ge/Mn(Ni)Bi$_2$Te$_4$.}
	\begin{ruledtabular}
		\begin{tabular}{cc}
			Building blocks & $T_{\rm C}$        \\ 
			\colrule
			Ge/MnBi$_2$Te$_4$ &  17.5K \\
			Ge/NiBi$_2$Te$_4$ &  18K  \\ 
			\colrule
			Ge/NiAs-Mn$_2$Bi$_2$Te$_5$ &  59K \\
			Ge/NaCl-Mn$_2$Bi$_2$Te$_5$ &  19K  \\
			Ge/NiAs-MnNiBi$_2$Te$_5$ &  182K \\
			Ge/NaCl-MnNiBi$_2$Te$_5$ &  119K  \\
			Ge/NiAs-NiMnBi$_2$Te$_5$ &  152K \\
			Ge/NaCl-NiMnBi$_2$Te$_5$ &  130K  \\
			Ge/NiAs-MnVBi$_2$Te$_5$ &  32K \\
			Ge/NaCl-MnVBi$_2$Te$_5$ &  80K  \\
			Ge/NiAs-NiVBi$_2$Te$_5$ &  108K \\
			Ge/NaCl-NiVBi$_2$Te$_5$ &  134K  \\	
		\end{tabular}
	\end{ruledtabular}
\end{table}	

Compared with that of Ge/Mn(Ni)Bi$_2$Te$_4$ whose $T_{\rm C}$ remain as low as about 18K, adopting $XY$Bi$_2$Te$_5$ commendably enhances the $T_{\rm C}$ above the boiling point of liquid nitrogen (77K) in most cases (except for only Ge/NiAs-MnVBi$_2$Te$_5$ with 32K), fully utilize the high temperature potential with the valley gaps analyzed in Sec. \ref{v:stacking-orders}.
	
Figures \ref{fig:figure13} (c)-\ref{fig:figure13}(f) serially depict the mapping distributions of $T_{\rm C}$ under stacking-order shifts on Ge/NiAs-Mn$_2$Bi$_2$Te$_5$, Ge/NiAs-NiMnBi$_2$Te$_5$, Ge/NiAs-MnVBi$_2$Te$_5$ and Ge/NaCl-NiVBi$_2$Te$_5$. In these mapping distributions, the colors from blue to white, yellow, then to red, the value of $T_{\rm C}$ increases. Distinctly, for Ge/NiAs-Mn$_2$Bi$_2$Te$_5$ and Ge/NiAs-NiMnBi$_2$Te$_5$, the $T_{\rm C}$ performs insensitively to the stacking orders, for the former derives from 60K to 70K, and the latter falls in the range 160K$\sim$170K. In Ge/NiAs-MnVBi$_2$Te$_5$, in most area $T_{\rm C}$ is fixed to about 31K while at (1/3, 0) and (0, 2/3) points it grows to about 60K. $T_{\rm C}$ is sensitive to stacking orders under the case of Ge/NaCl-NiVBi$_2$Te$_5$, within which the $T_{\rm C}$ value is divided into two major zones, including lower zone about 55K, and higher zone about 136K. Detailed analysis confirms that this large divergence is not originated from the interatomic-layer magnetic coupling reformations [Fig. S18(g)], nor from the MAE's deviations [Fig. S18(h)], but excellently fits well with first and second order of interatomic-layer exchange coefficients distributed in the stacking-order-shift zone [Figs. S19(g) and S19(h)]. Obviously, the reformed distribution between the first and second order of interatomic exchange coefficients act the key role to the final $T_{\rm C}$, in which the larger second-order coefficient, the higher interatomic-layer long-range magnetic couplings, and then the higher value of $T_{\rm C}$. This rule is also valid in the former three aforementioned conditions [comparing Fig. \ref{fig:figure13}(d)-\ref{fig:figure13}(f) with Figs. S19(a)-S19(f)].

As a summary, adopting $XY$Bi$_2$Te$_5$ inherits and exploits their high $T_{\rm C}$ natures, indicating that the tunability of multiple valley-based topology has the potential working under high temperatures ($\textgreater 130$K in some candidates). If the stacking-order shift is capable of revising the distribution between the first and second order of interatomic-layer exchange coefficients, the values of $T_{\rm C}$ are highly sensitive and tunable to the stacking orders.
	
	 \label{vi:tc}
	
\section{Conclusions and Guidance}
In this work, by constructing Xene/$XY$Bi$_2$Te$_5$ (Xene = germanene, silicene, or stanene) heterostructures, in conjunction with magnetic substrates such as Mn(Ni)Bi$_2$Te$_4$, NiAs-Mn$_3$Bi$_2$Te$_6$ and monolayer CGT, we have comprehensively investigated the manipulation of versatile valley-based magnetic topologies by solely modulating the magnetic properties of the underlying substrates. Based on the findings presented here, we propose a fully magnetic-substrate-driven, designable, and modulable approach for controlling various novel valley-based topological features.

\subsection*{Valley-polarized QAHE Relying Chiefly on the Nearest Magnetic Atomic Layer}
If the nearest magnetic atomic layer to the Xene exhibits intralayer FM coupling, it dominates the final valley-based topological character. Therefore, in most cases, magnetic vpQAHE state, operating within a wide range of valley-based magnetic topologies. The second-nearest magnetic atomic layer has a weak influence on the final valley-based topological character in most cases, except in the Ge/NiAs-MnNiBi$_2$Te$_5$ system, where FM and AFM couplings between the two magnetic layers induce vpQAHE and QAHE states, respectively.

\subsection*{Substrate-SOC Induced Manipulation}
The SOC strength of the magnetic substrate plays a crucial role in determining the final topological characters. By increasing the substrate-SOC strength, multiple topological phase transition points emerge, each corresponding to a different Chern number, with each transition point being associated with the closure of one of the valley gaps. Under certain conditions, a QVHE-QAHE phase transition occurs. Therefore, to obtain valley-based QAHE (QVHE) states, it is necessary to select a magnetic substrate with strong (weak) SOC strength or, accordingly, with heavy (light) element components.

\subsection*{Manipulations Based on In-plane Magnetic Orientations}
The in-plane magnetism generated by the magnetic substrate can also impart valley-based Chern insulating characters in Xene (taking germanene as a typical example). Both the value and chirality of the Chern numbers can be controlled by the magnetic moment orientations of the substrate. A moderate substrate-SOC strength is required for the above-mentioned in-plane magnetic orientation manipulation of valley-based topologies. If the substrate-SOC strength is too weak or too strong, this tunability is lost, leaving a Chern number of $C = 0$ across all magnetic orientations.

\subsection*{Valley-based Topology Manipulated via Stacking-order Shifts}
In Mn-neared systems (except for the FM state Ge/NiAs-MnNiBi$_2$Te$_5$), shifts in the stacking order can continuously adjust the values of the two valley gaps but only weakly affect the tunability of the Chern number, leaving a very narrow nonzero Chern insulating region. For Ni-neared building-blocks, however, the valley gaps, as well as the values and chirality of the Chern numbers, can be freely manipulated.

\subsection*{High Magnetic Critical Temperatures Rooted in the Magnetic Substrate}
High-temperature valley-based magnetic topologies can be established by selecting magnetic substrates with high magnetic critical temperatures, such as $XY$Bi$_2$Te$_5$. Notably, it is easier to select magnetic substrates with AFM coupling for high critical temperatures only if the nearest magnetic atomic layer maintains the FM state. This insight aids in the development of high-temperature, modulable valley-based magnetic topologies in practical applications.

The above guidance provides a complete roadmap for understanding and mastering the tunable factors behind the diverse valley-based magnetic topologies, offering experimentally executable and constructive references for future investigations. It is worth mentioning that one recent work demonstrated that stanene can be successfully grown on MnBi$_2$Te$_4$-family materials via molecular beam epitaxy \cite{barman2024growth}. Combined with well-documented achievements in the synthesis of silicene and germanene \cite{davila2014germanene,fleurence2012experimental,Jabra2022,Vogt2012,Lalmi2010,Meng2013,Li2014,Acapito2016,Ogikubo2020,Lin2012}, we are optimistic that the experimental fabrication of monolayer carbon-family films on various magnetic substrates will be achieved in the near future. Our study opens new avenues for exploring topological quantum states and their applications in robust, low-dissipation electronics, spintronics, and valleytronics, laying the groundwork for the next generation of functional quantum materials.

 \label{vii:guidance}
	
\begin{acknowledgments}
	
	We thank Prof. Ke He and Prof. Xiao Feng for financial supports, and also thank Prof. Ke He, Y. Chen and L. Y. Li for helpful discussions. This work was supported by the National Natural Science Foundation of China (Grant No. 92065206). Part of the numerical calculations has been done on the supercomputing system in the Huairou Materials Genome Platform.
	
	X. Y. Hong and Z. Li contributed equally to this work.
	
\end{acknowledgments}
\newpage
\nocite{*}
\bibliography{Main_text}

\end{document}